\documentclass[namedreferences]{SolarPhysics}
\usepackage[optionalrh]{spr-sola-addons} 
\usepackage{graphicx}                    
\usepackage{color}                       
\usepackage{url}                         

\usepackage[urlcolor=blue]{hyperref} 
\ifx \doiurl \undefined \def \doiurl#1{\href{http://dx.doi.org/#1}{\url{#1}}}\fi 
\ifx \adsurl \undefined \def \adsurl#1{\href{http://adsabs.harvard.edu/abs/#1}{\url{#1}}}\fi
\ifx \lrspurl \undefined \def \lrspurl#1{\href{http://www.livingreviews.org/#1}{\url{#1}}}\fi

\newcommand{\etal}{{\it et al.}}

\newcommand{\aap}{    {\it Astron. Astrophys.}}

\newcommand{\araa}{    {\it Ann. Rev. Astron. Astrophys.}}
\newcommand{\arfm}{    {\it Ann. Rev. Fluid Mech.}}
 
\newcommand{\apj}{    {\it Astrophys. J.}}

\newcommand{\gfd}{    {\it Geophys. Astrophys. Fluid Dyn.}}
\newcommand{\lrsp}{    {\it Living Rev. Solar Phys.}}

\newcommand{\nat}{    {\it Nature}}

\newcommand{\sci}{    {\it Science}}  
\newcommand{\solphys}{    {\it Solar Phys.}}

\providecommand{\e}[1]{\ensuremath{\times 10^{#1}}}

\begin{document}

\begin{article}

\begin{opening}

\title{Comparing Simulations of Rising Flux Tubes Through the Solar Convection Zone with Observations of Solar Active Regions: Constraining the Dynamo Field Strength}

%
\author{M.A.~\surname{Weber}$^{1,2}$\sep
        Y.~\surname{Fan}$^{1}$\sep
        M.S.~\surname{Miesch}$^{1}$      
       }

%
\runningauthor{M.A. Weber $\it{et \ al.}$}
\runningtitle{Comparing Simulations of Rising Flux Tubes with Observations of Active Regions}

%
  \institute{$^{1}$ High Altitude Observatory, National Center for Atmospheric Research, Boulder, CO, USA
                     email: \href{mailto:mariaw@ucar.edu}{mariaw@ucar.edu}  \\
                   $^{2}$ Colorado State University, Fort Collins, CO, USA
                    email: \href{mailto:mariaw@lamar.colostate.edu}{mariaw@lamar.colostate.edu}\\
             }

\begin{abstract}
We study how active-region-scale flux tubes rise buoyantly from the base of the convection zone to near the solar surface by embedding a thin flux tube model in a rotating spherical shell of solar-like turbulent convection.  These toroidal flux tubes that we simulate range in magnetic field strength from 15 kG to 100 kG at initial latitudes of 1$^{\circ} $ to 40$^{\circ}$ in both hemispheres.  This article expands upon Weber, Fan, and Miesch (\textit{Astrophys. J.}, 741, 11, 2011) (Article 1) with the inclusion of tubes with magnetic flux of $10^{20}$ Mx and $10^{21}$ Mx, and more simulations of the previously investigated case of $10^{22}$ Mx, sampling more convective flows than the previous article, greatly improving statistics. Observed properties of active regions are compared to properties of the simulated emerging flux tubes, including: the tilt of active regions in accordance with Joy's Law as in Article 1, and in addition the scatter of tilt angles about the Joy's Law trend, the most commonly occurring tilt angle, the rotation rate of the emerging loops with respect to the surrounding plasma, and the nature of the magnetic field at the flux tube apex.  We discuss how these diagnostic properties constrain the initial field strength of the active region flux tubes at the bottom of the solar convection zone, and suggest that flux tubes of initial magnetic field strengths of $\ge$ 40 kG are good candidates for the progenitors of large ($10^{21}$ Mx to $10^{22}$ Mx) solar active regions, which agrees with the results from Article 1 for flux tubes of $10^{22}$ Mx.  With the addition of more magnetic flux values and more simulations, we find that for all magnetic field strengths, the emerging tubes show a positive Joy's Law trend, and that this trend does not show a statistically significant dependence on the magnetic flux. 
\end{abstract}

%
\keywords{Interior, Convection Zone; Magnetic Fields, Models}

\end{opening}

%

\section{Introduction}

The toroidal magnetic field responsible for the emergence of solar active regions is believed to be generated by a dynamo mechanism at or near the base of the convection zone ({\it{e.g.}} \opencite{spiegel80}; \opencite{ball82}; \opencite{mi92}; \opencite{gilman00}; \opencite{char10}).  Of essential importance in the understanding of solar dynamo theory, and indeed all of solar physics, since flux emergence also regulates solar variability, is addressing how active region flux tubes rise dynamically through a turbulent convection zone. Also of interest is identifying the dynamo-generated magnetic field strength at the base of the convection zone required to produce the observed properties of solar active regions.      
 
Valuable insights into the nature and evolution of rising magnetic loops in the solar convective envelope have been gained through the use of the thin flux tube approximation by a plethora of authors ({\it{e.g.}} \opencite{spruit81}; \opencite{mi86}; \opencite{ferriz93}; \opencite{longcope97}; \opencite{fan09}).  It is found that in order for these simulated flux tubes to exhibit tilt angles and latitudes of emergence that agree well with observed solar active regions, the toroidal magnetic field at the base of the convection zone needs to be in the range of  $\approx$30 kG to $\approx$100 kG (\opencite{choud87}; \opencite{sch94}; \opencite{dsilva93}; \opencite{cali95}).  These studies also indicate that the Coriolis force is responsible for several observed asymmetries that exist between the leading (in the direction of solar rotation) and following polarities of solar active regions, such as: the tilt angles in accordance with Joy's Law (\opencite{dsilva93}; \opencite{cali95}), the apparent faster proper motion of the leading polarity of an emerging active region on the solar surface (\opencite{vandriel90}; \opencite{mi94}; \opencite{cali95}), and an asymmetry where the leading polarity shows a more coherent morphology (\opencite{fan93}; \opencite{cali95}; \opencite{cali98}).  However, as these previous studies neglect the effects of turbulent convection on rising flux tubes, it is possible that convective downdrafts can pin portions of the flux tube down to the base of the convection zone, while helical upflows between the downdrafts can boost the rise of the flux tube (\opencite{fan03}; \opencite{weber2011}).  Therefore, resulting emerging loops may exhibit different properties from those produced in the absence of convection and pose an intriguing topic of study.
 
As of yet, the magnetic field strength at which the solar dynamo operates is not well known, nor is it directly accessible via observations.  However, solar cycle dynamo models that incorporate the Lorentz force from large scale mean fields indicate that the magnetic field strength generated and amplified at the base of the convection zone is $\approx$15 kG, and most likely cannot exceed 30 kG (\opencite{rempel06a}; \opencite{rempel06b}).  Recent simulations of solar-like stars that rotate three times the current solar rate have shown that a rotating convective envelope can generate a dynamo that consists of opposite polarity magnetic wreaths in two hemispheres, which span the depth of the convection zone (\opencite{brown2010}).  When portions of these wreaths become strong enough, $\approx$35 kG or greater, a buoyant magnetic loop develops which then rises through the convecting fluid in which it is embedded (\opencite{nelson2011}).  While these dynamo-producing convection simulations are not meant to reproduce the solar dynamo directly, they do demonstrate that persistent toroidal fields can coexist with convection at moderate magnetic field strengths.  In light of these studies, it is important to understand how toroidal flux tubes of weak to moderate field strengths, $\approx$15 kG to $\approx$50 kG, behave as they rise through a turbulent solar convective envelope.  Some studies have been performed that investigate the buoyant rise of fully three-dimensional isolated flux tubes in a turbulent convective velocity field ({\it{e.g.}} \opencite{fan03}; \opencite{jouve09}).  However, due in part to the limited numerical resolution of these simulations, large values of magnetic flux must be used, which are greater than typical active region flux, in order to keep the tube from dissipating as it rises.

Recently, \opencite{weber2011} (hereafter Article 1) incorporate a thin flux tube model in a (separately computed) rotating three-dimensional convective velocity field representative of the solar convective envelope, in an effort to study the effects of turbulent solar-like convection on the dynamic evolution of active region scale emerging flux tubes.  Although the thin flux tube model cannot capture the possible fragmentation of the flux tube and its internal magnetic structure, it does preserve the frozen-in condition of the magnetic field.  The thin flux tube model is also useful in that each simulation can be performed quickly on single processor desktops, as compared to the three-dimensional simulations, which require multi-million processor hours on massively parallel supercomputers.  As such, the thin flux tube approach is a useful platform for a parameter space study, and provides a starting point for the investigation of the effects of convective flows on flux tubes of realistic magnetic field strengths with magnetic fluxes similar to those of active regions observed on the Sun.

This article builds upon the work as presented in Article 1.  In Article 1, it was found that as the magnetic field strength increases from 15 kG to 100 kG, the dynamic evolution of the emerging flux loops changes from being convection dominated to magnetic buoyancy dominated.  For these flux tubes, the convective flow is found to reduce the rise time and the latitude of emergence through the anchoring of flux loop footpoints by downdrafts.  The addition of solar-like convection also promotes tilt angles that are consistent with the observed mean tilt of solar active regions in part because of the mean kinetic helicity of the upflows in the simulated convection. In this article, we expand the study by increasing the number of simulations performed by about 3.5 times per magnetic field strength to improve the statistics of the results, sampling different time spans of the convective flows.  Solar active regions exhibit magnetic flux in the range of $10^{20}$ Mx to $10^{22}$ Mx, which include ephemeral regions and pores to the largest scale sunspots (\opencite{zwaan87}).  To study the dependence of flux tube evolution on magnetic flux, we also include cases with flux of $10^{20}$ Mx and $10^{21}$ Mx in this study in addition to the $10^{22}$ Mx cases. This in combination with the increased number of simulations per magnetic field strength results in a total of 5940 flux tubes to analyze.  As with Article 1, we consider flux tubes with initial latitudes of 1$^{\circ}$ to 40$^{\circ}$ in both the northern and southern hemispheres. Besides the observed active region properties described by Joy's Law, in this article we also consider additional observational diagnostics, including the scatter of active region tilts and sunspot rotation rates to further constrain the initial field strength of active region flux tubes at the base of the solar convection zone.  We also discuss the properties of the magnetic field at the apices of the emerging flux loops.  In comparison to Article 1, our new study confirms that with convection all flux tubes with magnetic field strengths of 15 kG -- 100 kG do exhibit a positive Joy's Law trend, and given the uncertainties that we obtain, we note no significant dependence of the flux tube tilt angle on magnetic flux.

A description of the thin flux tube model and how it is incorporated with the spherical shell of solar-like convection employed in this study is outlined in Section \ref{sec:model}.  We highlight the results obtained from our simulations in Section \ref{sec:results}, discussing flux tube rise times, latitudes of emergence, tilt angles, magnetic field properties, and the rotation rate of the simulated emerging loops.  A summary of the results is given in Section \ref{sec:discuss}. 

\section{Model Description}
\label{sec:model}

The dynamics of a thin, isolated magnetic flux tube can be described by the thin flux tube approximation.  These equations are derived from the ideal MHD equations, operating under the assumption that the flux tube radius is small compared to all other relevant scales of variation associated with the flux tube ({\it{e.g.}} \opencite{spruit81}; \opencite{cheng92}; \opencite{longcope97}; \opencite{fan09}). As in Article 1, we solve the following equations, which describe the evolution of each Lagrangian element of the one-dimensional flux tube:

\begin{eqnarray}
\rho {d {\bf v} \over dt} & = & -2 \rho ( {\bf \Omega_0} \times {\bf v} )
-(\rho_e - \rho ) [{\bf g} - {\bf \Omega_0} \times ({\bf \Omega_0} \times
{\bf r})] + {\bf l} {\partial \over \partial s} \left ( { B^2 \over 8 \pi}
\right ) + {B^2 \over 4 \pi} {\bf k} 
\nonumber \\
& & - C_d {\rho_e | ({\bf v }-{\bf v}_e)_{\perp} |
({\bf v}-{\bf v}_e)_{\perp} \over ( \pi \Phi / B )^{1/2} }, 
\label{eq:eqn_motion}
\end{eqnarray}
\begin{equation}
{d \over dt} \left ( {B \over \rho} \right ) = {B \over \rho} \left [
{\partial ({\bf v} \cdot {\bf l}) \over \partial s} - {\bf v} \cdot
{\bf k} \right ] , 
\label{eqn_cont_induc}
\end{equation}
\begin{equation}
{1 \over \rho} {d \rho \over dt} = {1 \over \gamma p} {dp \over dt} ,
\label{eqn_adiab}
\end{equation}
\begin{equation}
p = {\rho R T \over \mu} ,
\label{eqn_state}
\end{equation}
\begin{equation}
p + {B^2 \over 8 \pi} = p_e ,
\label{eqn_pbalance}
\end{equation} 
where, ${\bf r}$, ${\bf v}$, {\it{B}}, {\it{$\rho$}}, {\it{p}}, {\it{T}}, which are functions of time {\it{t}} and arc length {\it{s}} measured along the tube, denote respectively the position, velocity, magnetic field strength, gas density, pressure, and temperature of a Lagrangian tube segment with subscript $e$ referring to the external plasma which are functions of depth only, ${\bf l} \equiv \partial {\bf r} / \partial s$ is the unit vector tangential to the flux tube, and ${\bf k} \equiv \partial^2 {\bf r} / \partial s^2 $ is the tube's curvature vector, subscript ${\perp}$ denotes the component perpendicular to the flux tube, $\Phi$ is the constant total flux of the tube, $\mu$ is the mean molecular weight of the surrounding external plasma, ${\bf g}$ is the gravitational acceleration and a function of depth, ${\bf \Omega_0}$ is the angular velocity of the reference frame co-rotating with the sun, with $\Omega_0$ set to $2.7 \times 10^{-6}$ rad s$^{-1}$ in this calculation, $C_d$ is the drag coefficient set to unity, $\gamma$ is the ratio of specific heats, $R$ is the ideal gas constant, and ${\bf v}_e ({\bf r}, t)$ is a time-dependent velocity field relative to the rotating frame of reference that impacts the dynamics of the thin flux tube through the drag force term. In the above equations, we do not introduce an explicit magnetic diffusion or kinematic viscosity term.  However, we do introduce a drag force (last term in Equation (\ref{eq:eqn_motion})), which describes the interaction of the external fluid with the flux tube in a high Reynolds number regime ({\it{e.g.}} \opencite{batchelor}).  The flux tube evolves passively in the external fluid such that the tube imparts no back reaction on the fluid in which it is embedded. The magnetic field of the flux tube is untwisted such that it only has a component in the $\hat{l}$ direction, and the tube is discretized with 800 uniformly spaced grid points along its arc length {\it{s}}.  A description of the numerical methods used to solve the flux tube evolution as determined by the above set of equations is discussed in detail by \inlinecite{fan93}.  

The thin flux tube approximation is no longer satisfied when the radius of the flux tube is on the order of the local pressure scale height.  With this in mind, the thin flux tube approximation is satisfied for the bulk of the convection zone, where the flux tube spends the majority of its time during evolution. For large flux values of $\approx$$10^{22}$ Mx, and small magnetic field strength of $\approx$15 kG, the radius of the flux tube will be the greatest, and will expand significantly as it rises toward the top of the convection zone.  Due to these effects, we must stop our simulation before the flux tube can reach the photosphere. In this case, we stop our simulations once the flux tube has reached a height of $\approx$$21$ Mm below the solar surface.

A reference one-dimensional solar structure model \linebreak (\opencite{jcd1996}) provides the stratification parameters of the external field-free plasma in the convection zone, with an extension of a simple polytropic, sub-adiabatically stratified thin overshoot layer, as described by \inlinecite{fan_gong2000}.  The sub-adiabatic overshoot region extends from \linebreak 4.8\e{10} cm ($r=0.69 R_{\odot}$) to 5.026\e{10} cm from Sun center, with the super-adiabatic convection zone extending from 5.026\e{10} cm to 6.75\e{10} cm ($r=0.97 R_{\odot}$), which is still $\approx21$ Mm below the solar surface. Profiles of $T_e$, $\rho_e$, $p_e$, and the adiabaticity $\delta$, we use can be found in Article 1. 

What sets the thin flux tube simulations in Article 1 (and continued here) apart from previous simulations is the inclusion into the drag force term of a time-dependent convective velocity field ${\bf v}_e ({\bf r}, t)$ relative to the rotating frame of reference.  This three-dimensional  global convection simulation is computed separately from the thin flux tube simulation using the ASH (anelastic spherical harmonic) code, as described by \inlinecite{mieschetal2006} (hereafter MBT06).  In the anelastic approximation, the velocity of convection flows is taken to be much slower than the speed of sound in the fluid, and convective flows and thermal variations are treated as a linear perturbation to a background state taken from a one-dimensional solar structure model.  The computed ${\bf v}_e ({\bf r}, t)$ captures giant-cell convection and the associated mean flows such as meridional circulation and differential rotation, in a rotating convective envelope spanning $r = 0.69 R_{\odot}$ to $r = 0.97 R_{\odot}$ (4.8\e{10} cm to 6.75\e{10} cm from Sun center).  ASH is a pseudo-spectral code, and the particular simulation used here is resolved by a grid of 129 points in $r$, $256$ points in $\theta$, and $512$ points in $\phi$.  To find more specific details regarding this ASH simulation, see Article 1.    

Boundary conditions and simulation parameters for the ASH simulation are similar to Case AB3 in MBT06.   Particularly, the lower thermal boundary condition is the same as in Case AB3, with a latitudinal entropy gradient imposed to implicitly capture thermal coupling to the tachocline.  This feature helps promote a conical rotation profile (see MBT06).  The radial entropy gradient imposed at the outer boundary is steeper in this simulation, and therefore more in line with solar structure models ({\it{e.g.}} \opencite{jcd1996}), as compared to Case AB3.  In this simulation the values of the thermal diffusivity $\kappa$ and the turbulent viscosity $\nu$ at the outer boundary ($r=0.97R_{\odot}$) are $4\times 10^{13}$ cm$^2$ s$^{-1}$ and $2\times10^{13}$ respectively, and each decreases with depth in proportion to the inverse square root of the background density $\hat{\rho}^{-1/2}$.  Somewhat larger than the density contrast of Case AB3 due to a slightly deeper domain, the density contrast across this ASH simulation domain is 69, corresponding to 4.2 density scale heights.  This yields a mid-convection zone Rayleigh number $R_a$ of $5\times 10^6$ and Reynolds number $R_e$ of order 50.  The value of $R_{e}$ in this simulation is about a factor of two smaller than that used in the flux tube simulations of \inlinecite{jouve09}, but the value of $R_{a}$ is larger.

While this ASH simulation is more laminar than some others (\opencite{miesch2008}; \opencite{jouve09}), it still possesses all of the relevant features necessary to explore the fundamental interactions between thin flux tubes and the mean flows associated with global convection. These features include: asymmetric, rotationally aligned cells at low latitudes (density-stratified banana cells), rapidly-evolving downflows in the upper convection zone at high latitudes dominated by helical plumes, and a strong, solar-like differential rotation.  Even the highest resolution simulations exhibit similar basic features, therefore we would not expect the essential results to change significantly with more turbulent convection.

\begin{figure}
\centerline{\includegraphics[width=0.6\textwidth,clip=]{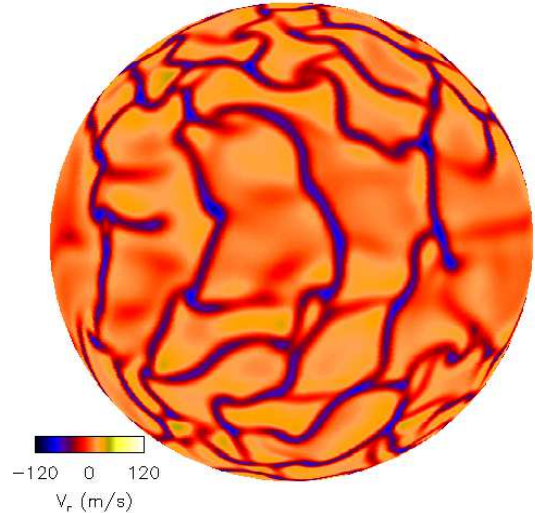}}
\caption{A snapshot of the convective radial velocity at a depth of 25 Mm below the solar surface.}
\label{fig:vr}
\end{figure}

A typical giant-cell convection pattern at a depth of 25 Mm below the solar surface is shown in Figure \ref{fig:vr}. Broad upflow cells are surrounded by narrow downflow lanes, which can reach maximum downflow speeds of nearly $600$ m s$^{-1}$ at a mid-convection zone depth of about $86$ Mm below the surface.  Throughout the bulk of the convection zone, the combined influence of density stratification and the Coriolis force induces anti-cyclonic vorticity in expanding upflows and cyclonic vorticity in contracting downflows.  These effects yield a mean kinetic helicity density $H_k$ that is negative in the northern hemisphere and positive in the southern hemisphere (see Article 1 for an image of the associated kinetic helicity).   Such a helicity pattern is typical for rotating, compressible convection ({\it{e.g.}} \opencite{miesch09}).

Columnar, elongated downflow lanes align preferentially with the rotation axis at low latitudes, reflecting the presence of so-called ``banana cells'' (see Figure \ref{fig:vr}). Such structures propagate in a prograde direction relative to the polar regions, due in part to differential rotation and an intrinsic phase drift similiar to traveling Rossby waves (\opencite{miesch09}).  The total angular velocity $\Omega/2 \pi$ (with respect to the inertial frame) is solar-like, and decreases monotonically from $\approx$470 nHz at the equator to $\approx$330 nHz at the poles, and exhibits nearly conical contours at mid-latitudes (see Article 1 for an image of the associated differential rotation), as observed in the solar convection zone (\opencite{thompson2003}).

As is in Article 1, our simulations start with isolated toroidal magnetic flux rings in mechanical equilibrium (neutral buoyancy), located at a radial distance to the center of the Sun of $r = r_0 = 5.05 \times 10^{10}$ cm, slightly above the base of the convection zone. To ensure initial neutral buoyancy, the internal temperature of the flux tube is reduced compared to the external temperature.  The toroidal ring is perturbed with small undular motions which consist of a superposition of Fourier modes with azimuthal order ranging from $m = 0$ through $m = 8$  with random phase relations. In this work and in Article 1, the external time dependent three-dimensional convective flow described impacts the flux tube through its drag force term.  We consider a range of initial field strengths of 15 kG, 30 kG, 40 kG, 50 kG, 60 kG, and 100 kG, and initial latitudes ranging from 1$^{\circ}$ to 40$^{\circ}$ for the toroidal flux ring, with a constant magnetic flux of $10^{20}$ Mx to $10^{22}$ Mx.  This range of magnetic flux values is typical of ephemeral regions and pores to the strongest sunspots.  We also greatly improve statistics by performing seven simulations sampling different time ranges of the ASH convective flow for each magnetic field strength, initial latitude, and flux, for a total number of flux tube simulations of 5940.  The thin flux tube simulations sample a span of $\approx$2.7 years of consecutive ASH convective flows.

\section{Results}
\label{sec:results}

\subsection{Rise Time and Latitude of Emergence}
\label{sec:dynamics}

\begin{figure}
 \centerline{\hspace*{0.015\textwidth}
               \includegraphics[width=0.515\textwidth,clip=]{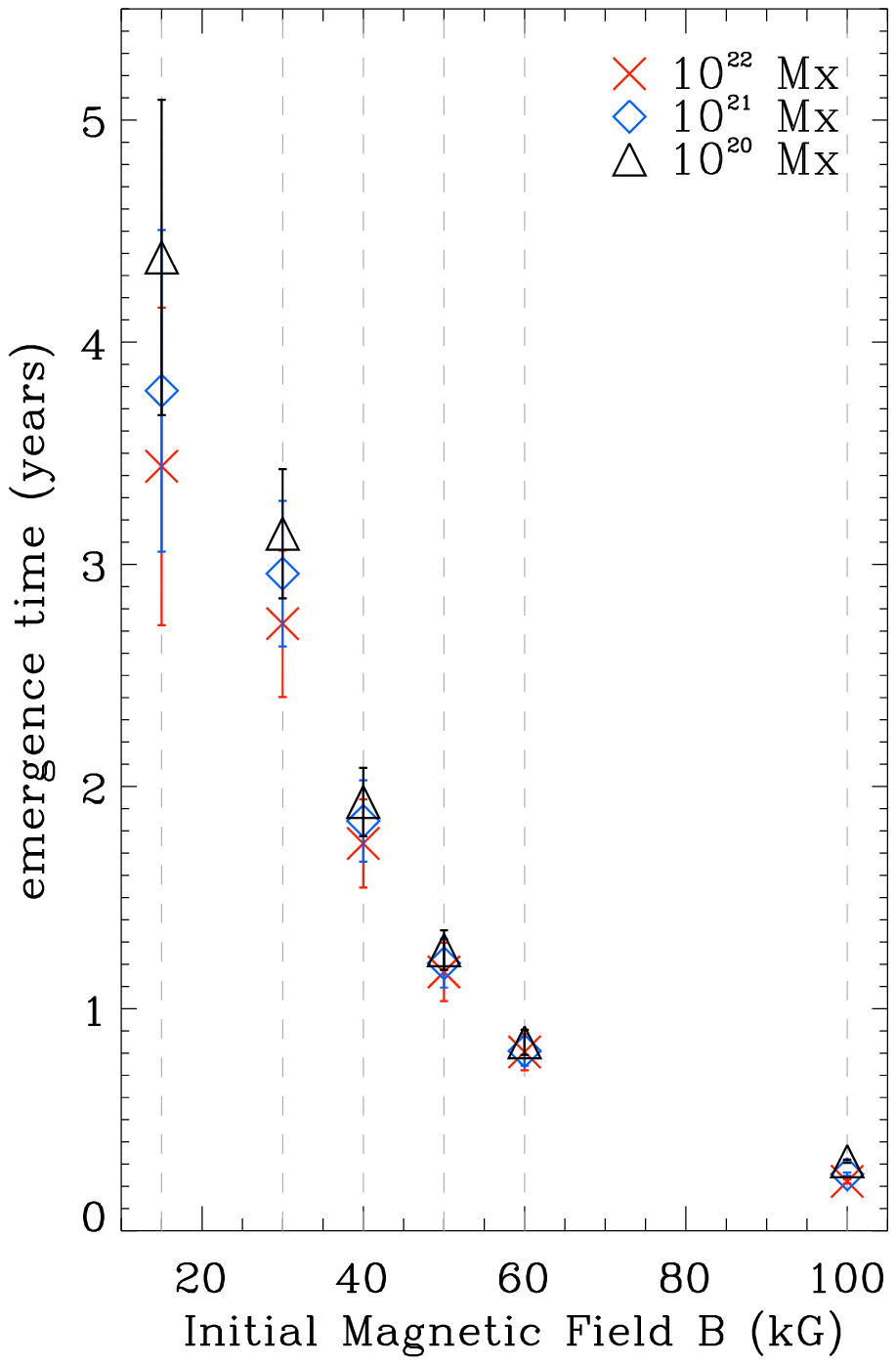}
               \hspace*{-0.03\textwidth}
               \includegraphics[width=0.515\textwidth,clip=]{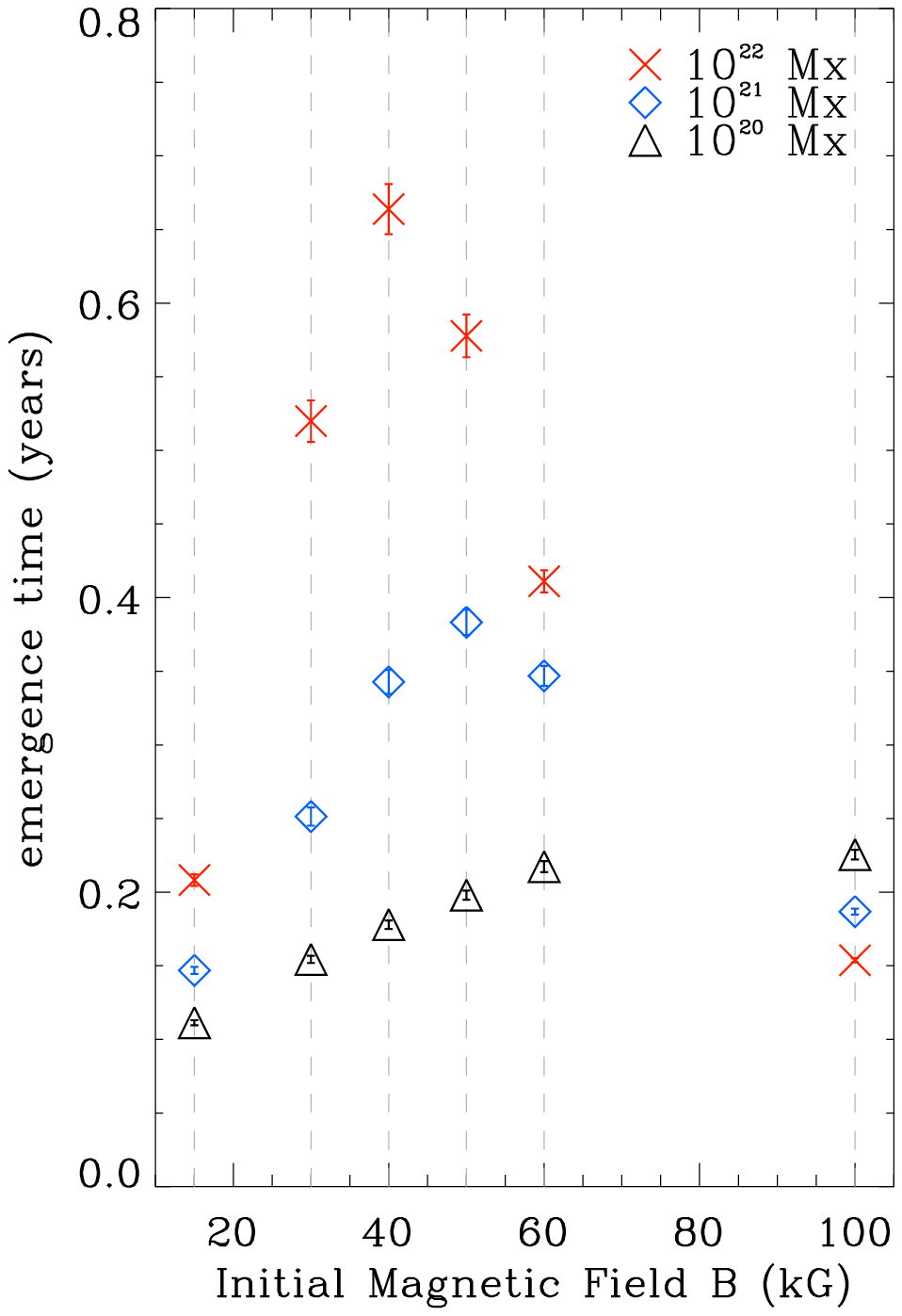}
              }
              
     \vspace{-0.7\textwidth}   
     \centerline{\large \bf     
      \hspace{0.05 \textwidth}  \color{black}{(a)}
      \hspace{0.43\textwidth}  \color{black}{(b)}
         \hfill}
     \vspace{0.66\textwidth}    
              
 \caption{(a) Average rise times for flux tube simulations without the influence of convection.  Red crosses are for flux tubes with magnetic flux of $10^{22}$ Mx, blue diamonds for $10^{21}$ Mx, and black triangles for $10^{20}$ Mx.  Each symbol represents an average of nine flux tube simulations. This figure shows a decrease in  emergence time with an increase in magnetic field strength and flux.  (b)  Average rise times for flux tube simulations with the influence of convection.  Each symbol represents an average of $\approx$330 flux tube simulations. With the addition of convection, rise times of the flux tubes are shorter for tubes with smaller magnetic flux, except at 100 kG where the trend is reversed. Flux tubes of mid-field strength now take the longest time to emerge. Bars represent the standard deviation of the mean.  }             
\label{fig:time}
\end{figure}

The drag force term (the last term in Equation (\ref{eq:eqn_motion})) depends upon the magnetic field $B$ and the flux $\Phi$ through the ratio of $(\Phi/B)^{1/2}$, which is proportional to the cross-sectional radius of the  thin flux tube.  As this ratio appears in the denominator of the drag force term, thinner tubes will experience more drag.  For any given magnetic field, the diameter of the flux tube can be reduced by decreasing the magnetic flux.  In the absence of convection, the drag force acting on the rising tube reduces the velocity of the flux tube in all directions, increasing its rise time and reducing its latitude of emergence.  Figure \ref{fig:time}-(a) shows a trend of increasing rise time of the flux tube simulations for both a decreasing magnetic field strength and a decreasing magnetic flux in the absence of convection, which has been found in previous thin flux tube simulations without the influence of convection ({\it{e.g.}} \opencite{mi83}; \opencite{choud87}; \opencite{dsilva93}; \opencite{fan93}).  Taking these effects into consideration, a tube with a reduced flux and a reduced magnetic field should take the longest to rise, as is shown in Figure \ref{fig:time}-(a), with 15 kG, $10^{20}$ Mx tubes taking the longest time to rise. 

However, with the addition of convection, flux tubes with smaller magnetic flux exhibit a shorter rise time, except for those with an initial magnetic field strength of 100 kG, as shown in Figure \ref{fig:time}-(b).  In this case, the drag force term affects how strongly the flux tube is coupled with the convective velocity field.  Rising flux tubes with lower magnetic flux are advected strongly by convection.  This aids the flux tube in emerging at the surface faster than it could at a larger magnetic flux, provided the drag force due to convection is significant compared to the buoyancy force.  On the other hand, the magnetic buoyancy and magnetic tension of flux tubes with an initial magnetic field strength of 100 kG dominate the drag force due to convection.  As a result, flux tubes of $10^{20}$ Mx still take a longer time to emerge than those with a flux of $10^{22}$ Mx, as is the trend without convection.  Similar to Article 1, flux tubes with mid-field strengths of 40 kG\,--\,50 kG and magnetic flux of $10^{21}$ Mx and $10^{22}$ Mx take the longest time to emerge.  For these magnetic flux and magnetic field strengths, the average convective downflows and magnetic buoyancy of the flux tube are of similar magnitudes (see Article 1).  A $\it{tug-of-war}$ exists between these two effects until one eventually dominates.  Considering the $\approx$11-year duration of the solar cycle, for flux tubes originating near the convection zone base, a maximum ruse time of about eight months for flux tubes subject to convective effects (see Fig. \ref{fig:time}-(b)) is much more realistic than a maximum rise time of about five years without convection (see Fig. \ref{fig:time}-(a)).

\begin{figure}
\centerline{\includegraphics[width=0.6\textwidth,clip=]{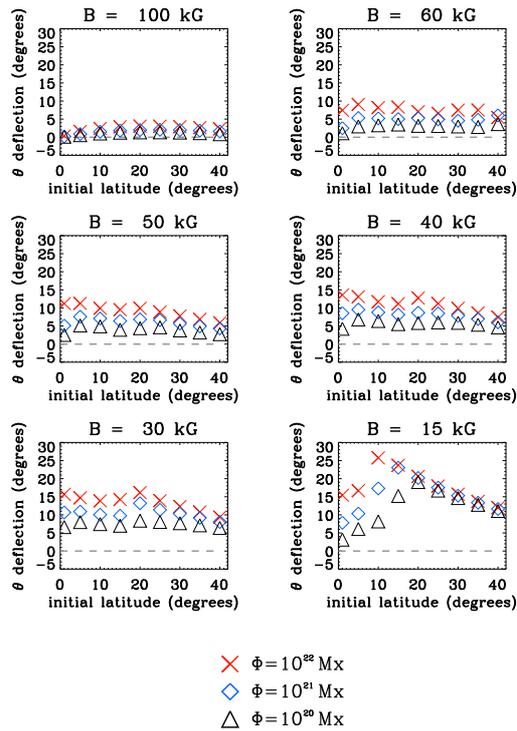}}
\vspace{-0.07\textwidth}    
\caption{Latitudinal deflection (emergence latitude minus initial latitude) of the flux tube apex as a function of initial latitude for initial flux tube magnetic field strengths of 100 kG, 60 kG, 50 kG, 40 kG, 30 kG, and 15 kG in the absence of convection for magnetic flux of $10^{22}$ Mx (red crosses), $10^{21}$ Mx (blue diamonds), and $10^{20}$ Mx (black triangles).  An increased drag force for tubes with a smaller flux reduces the poleward deflection of the tube.}
\label{fig:emlat}
\end{figure}

Without the external flow field, the increased drag force on flux tubes at lower magnetic flux aids in reducing the poleward deflection of the flux tube by the Coriolis force such that the tube rises more radially.  As a result, the flux tube emerges at lower latitudes for lower magnetic flux, as has been found in previous thin flux tube calculations in the absence of convection \linebreak ({\it{e.g.}} \opencite{choud87}; \opencite{dsilva93}; \linebreak \opencite{fan93}). This is supported by Figure \ref{fig:emlat} which shows the latitudinal deflection (emergence latitude minus initial latitude) of the flux tube as a function of initial latitude in the absence of convection for magnetic flux of $10^{20}$ Mx (black triangles), $10^{21}$ Mx (blue diamonds), and $10^{22}$ Mx (red crosses).  Also, as found in previous studies (\opencite{choud87}; \opencite{cali95}), a low-latitude zone with the absence of flux emergence due to poleward deflection is found for magnetic field strengths of 15 kG to 40 kG.  As such, these flux tubes would not be able to produce active regions near the equator.

\begin{figure}
\centerline{\includegraphics[width=0.95\textwidth,clip=]{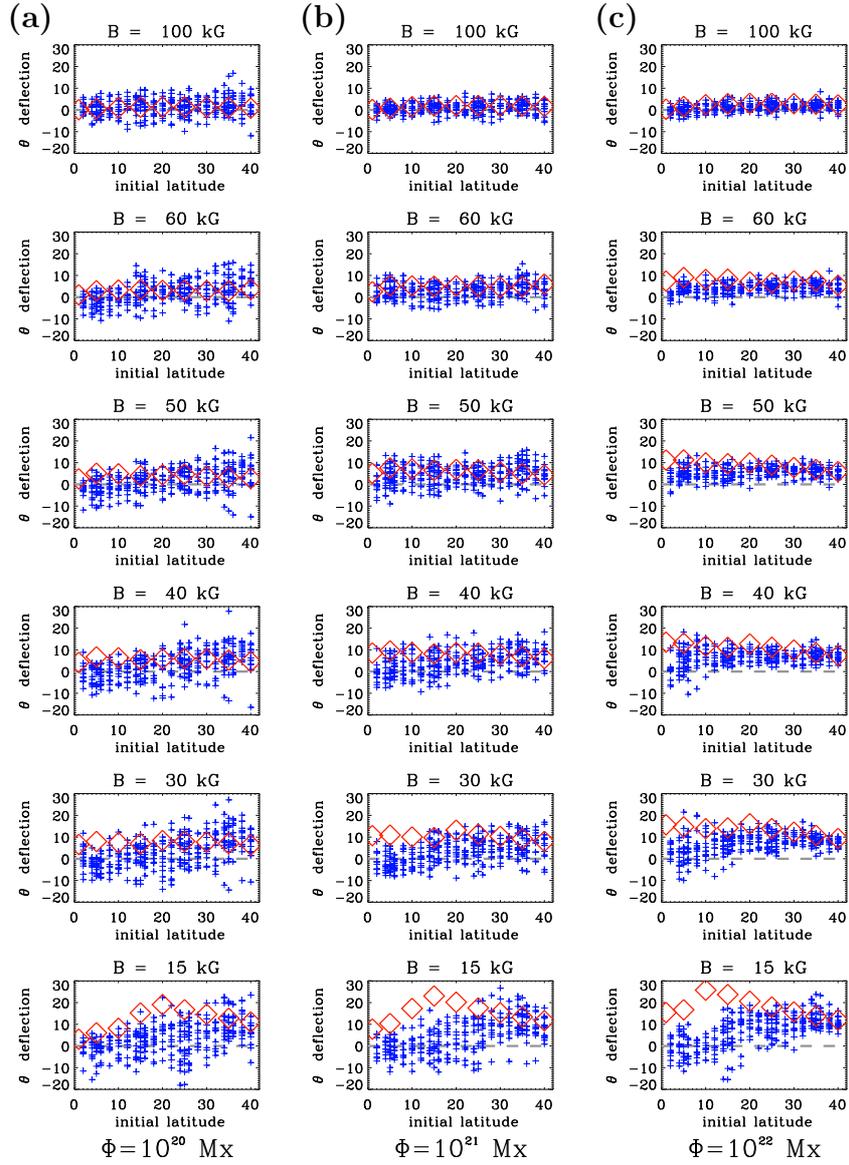}}
              \vspace{-1.45\textwidth}   
     \centerline{\large \bf     
      \hspace{0.03 \textwidth}  \color{black}{(a)}
      \hspace{0.24\textwidth}  \color{black}{(b)}
      \hspace{0.24\textwidth}  \color{black}{(c)}
       \hfill}
     \vspace{1.3\textwidth}    

\caption{Latitudinal deflection (emergence latitude minus initial latitude) of the flux tube apex as a function of initial latitude for initial flux tube magnetic field strengths of 100 kG, 60 kG, 50 kG, 40 kG, 30 kG, and 15 kG with a magnetic flux of (a) $10^{20}$ Mx in column 1, (b) $10^{21}$ Mx in column 2, (c) and $10^{22}$ Mx in column 3.  Red diamond are for flux tubes without convection, and plus signs are for flux tubes with convective effects. Both axes are in units of degrees.  Addition of a convective velocity field results in flux tubes that are able to emerge near the equator at lower magnetic fields.}
\label{fig:emergencelat}
\end{figure}

With convection, the previous problem of poleward slippage for flux tubes of weak initial field strength is rectified.  Convection produces a scatter in the emerging latitude compared to the cases without convection, which is both poleward and equatorward. Figure \ref{fig:emergencelat} shows the latitudinal deflection of the flux tube for fluxes of $10^{20}$ Mx, $10^{21}$ Mx, and $10^{22}$ Mx resulting from simulations with convection (plus signs), compared to the case without convection (diamonds). A negative latitudinal deflection indicates that the flux tube emerges close to the equator than where it originally started.  Such a phenomenon only occurs due to external convective flows.  It is even possible for portions of the flux tube to cross into the opposite hemisphere.  We note a systematic trend of reduced poleward deflection of the flux tube for all magnetic fluxes considered here.  This is in agreement with Article 1, where we only studied $10^{22}$ Mx cases.  It is also evident that there is a greater amount of scatter in latitudinal deflection at lower magnetic flux, which occurs because the tube is advected more by convective flows than at a larger magnetic flux.  Also, as the magnetic field strength increases for a particular flux, the latitudinal deflection decreases, as is the case for tubes both with and without convection.  This is a result of the buoyancy force overpowering the Coriolis force at large magnetic fields, forcing the tube to rise more radially.         

\subsection{Joy's Law} 
\label{sec:joy}
The phenomenon associated with solar active regions known as Joy's Law describes the tilting behavior of emerging flux regions toward the equator ({\it{e.g.}} \opencite{hale19}).  A line drawn between the center of the two bipolar regions will be tilted with respect to the east-west direction such that the leading polarity (in the direction of solar rotation) is closer to the equator than the following.  On average, the tilt of this line with respect to the East\,--\,West direction, quantified by the tilt angle, increases with an active region's latitude of emergence.  In this article, the tilt angle is computed as the angle between the tangent vector at the apex of the emerging loop (once it has reached the top of the simulation domain), and the local East\,--\,West direction.  We define a positive sign of tilt as a clockwise (counter-clockwise) rotation of the tangent vector away from the East\,--\,West direction in the northern (southern) hemisphere, consistent with the direction of the observed mean tilt of active regions.  If the magnitude of the tilt angle exceeds 90$^{\circ}$, then the active region is of an anti-Hale arrangement, possessing the wrong leading polarity in the direction of solar rotation for that particular hemisphere.  This is the same convention as used in Article 1.   

\begin{figure}
\centerline{\includegraphics[width=0.95\textwidth,clip=]{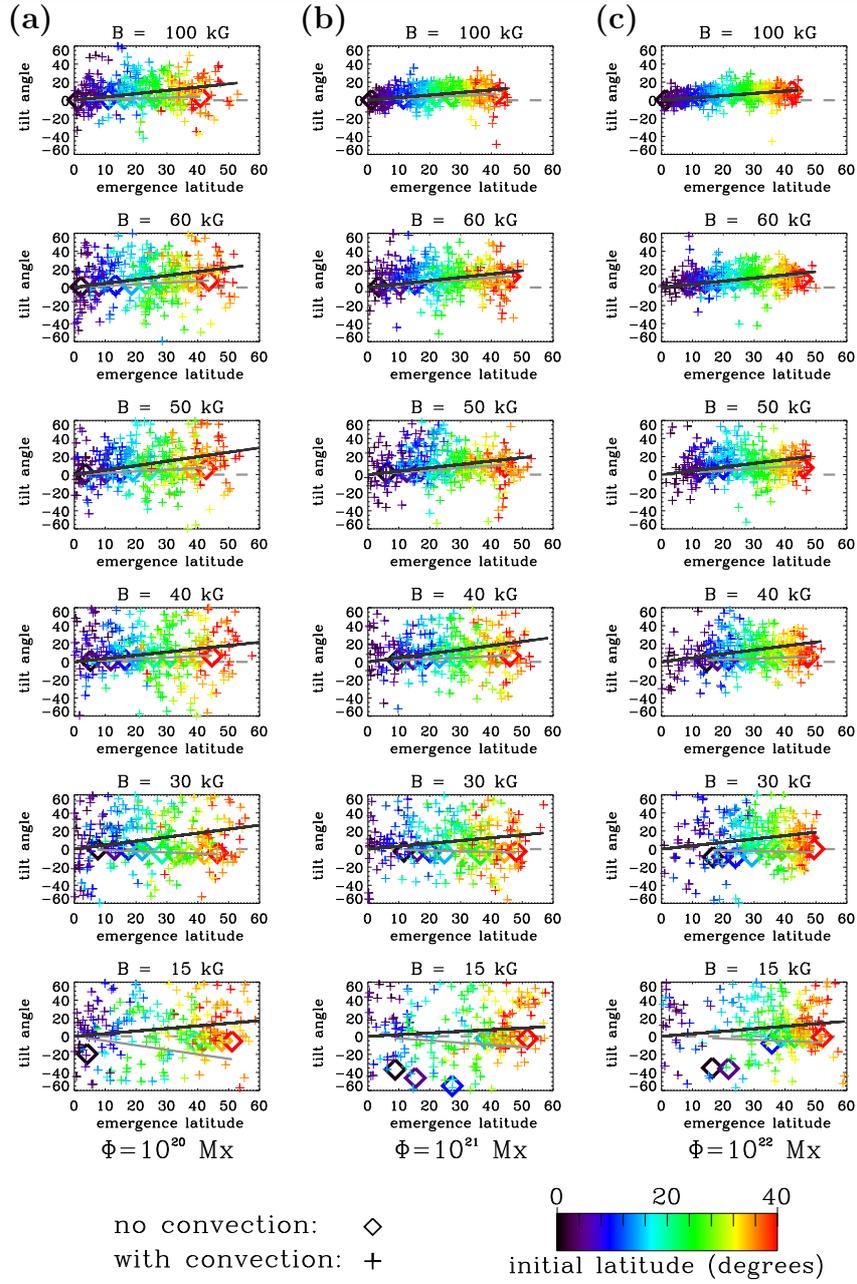}}
              \vspace{-1.45\textwidth}   
     \centerline{\large \bf     
      \hspace{0.03 \textwidth}  \color{black}{(a)}
      \hspace{0.24\textwidth}  \color{black}{(b)}
      \hspace{0.24\textwidth}  \color{black}{(c)}
       \hfill}
     \vspace{1.45\textwidth}    

\caption{Tilt angles as a function of emergence latitude for initial magnetic field strengths of 100 kG, 60 kG, 50 kG, 40 kG, 30 kG, and 15 kG for cases with (plus signs) and without (diamond points) the influence of convection for magnetic flux of (a) $10^{20}$ Mx in column 1, (b) $10^{21}$ Mx in column 2, and (c) $10^{22}$ Mx in column 3, with data sets sampling seven different convective velocity flow fields.  The gray line is the linear best fit for flux tubes in the absence of convection, and the black line is the best-fit line for the cases subjected to convective flows. Slopes for all of these best-fit lines are shown in Table 1. A color bar indicates the original starting latitude of the flux tube.  All axes are in units of degrees.  Convection introduces a scatter of the tilt angle about the best-fit line, and also aids in increasing the slope of the best-fit line, especially at lower magnetic field strengths.}
\label{fig:tiltangle}
\end{figure}

Tilt angles of the flux tube as a function of emergence latitude for magnetic flux of $10^{20}$ Mx, $10^{21}$ Mx, and $10^{22}$ Mx are shown in Figure \ref{fig:tiltangle}, for initial magnetic field strengths of 100 kG, 60 kG, 50 kG, 40 kG, 30 kG, and 15 kG both with (plus signs) and without convection (diamonds). Each of the panels in this figure includes about 3.5 times more flux tube simulations per magnetic field strength compared to Article 1, sampling different time ranges of the ASH convective velocity flow, therefore greatly improving statistics.  As a result, each panel contains $\approx$330 points. Here we also expand the investigation to include cases with flux of $10^{20}$ Mx and $10^{21}$ Mx. As in Article 1, we assume that the tilt angle increases monotonically with increasing emergence latitude, and so perform a linear least-squares fit of Joy's Law to the data using the equation: $\alpha = m_{A} \lambda$, where $\alpha$, $m_{A}$, and $\lambda$ represent the tilt angle, slope, and latitude respectively.  The fit is forced to go through zero because no tilt is expected for equatorial sunspot groups.  Figure \ref{fig:tiltangle} shows in gray the best-fit line for the data without convection, and black for the flux tubes with convection.  The slopes of these best-fit lines along with their uncertainties are reported in Table \ref{table1} for flux tubes without convective effects, and Table \ref{table2} for flux tubes with convective effects.  

\begin{table}
\begin{tabular}{cccc}
\hline
\emph{B} & $10^{20}$ Mx & $10^{21}$ Mx & $10^{22}$ Mx \\
 without convection \\
\hline
100 kG &  0.10 $\pm$ 0.01   & 0.16 $\pm$ 0.01      & 0.25 $\pm$ 0.01 \\
60 kG   &  0.18 $\pm$ 0.01    & 0.27 $\pm$ 0.01      &  0.30 $\pm$ 0.02 \\
50 kG   &  0.18 $\pm$ 0.01    & 0.25 $\pm$ 0.01      & 0.22 $\pm$ 0.02  \\
40 kG   &  0.16 $\pm$ 0.01    & 0.16 $\pm$ 0.01      & 0.12 $\pm$ 0.01  \\
30 kG   &  -0.12 $\pm$ 0.01  & -0.08 $\pm$ 0.03    & -0.07 $\pm$ 0.05  \\
15 kG   &  -0.50 $\pm$ 0.38    & -0.24 $\pm$ 0.22    & -0.13 $\pm$ 0.13   \\
\hline

\end{tabular}
\caption{Slopes $m_{A}$ of the linear best-fit lines to the tilt angle as a function of emergence latitude for flux tubes without convective effects.  At low magnetic field strengths, the slope of the best-fit line is negative, indicating a departure from the Joy's Law trend.}
\label{table1}
\end{table}

\begin{table}
\begin{tabular}{cccc}
\hline
\emph{B} & $10^{20}$ Mx & $10^{21}$ Mx & $10^{22}$ Mx \\
 with convection \\
\hline
100 kG &  0.36 $\pm$ 0.05   &  0.28 $\pm$ 0.02     & 0.25 $\pm$ 0.02\\
60 kG   &  0.44 $\pm$ 0.08    & 0.37 $\pm$ 0.04     & 0.35 $\pm$ 0.02\\
50 kG   &  0.50 $\pm$ 0.08    &  0.38 $\pm$ 0.04    & 0.42 $\pm$ 0.04\\
40 kG   &  0.36 $\pm$ 0.10    & 0.45 $\pm$ 0.08     & 0.44 $\pm$ 0.06\\
30 kG   &  0.44 $\pm$ 0.11    & 0.31 $\pm$ 0.11     & 0.37 $\pm$ 0.09 \\
15 kG   &  0.29 $\pm$ 0.12    & 0.18 $\pm$ 0.14     & 0.27 $\pm$ 0.12 \\
\hline

\end{tabular}
\caption{Slopes $m_{A}$ of the linear best-fit lines to the tilt angle as a function of emergence latitude for flux tubes with convective effects. Convection aids in increasing the slope of the best-fit line, and the slope also peaks at mid-field strengths of 40 kG and 50 kG.  In comparison, using white light sunspot group tilt angle data, Dasi-Espuig $\it{et}$ $\it{al.}$ (2010) find $m_{A}=0.26\pm0.05$ for Mount Wilson data and $m_{A}=0.28\pm0.06$ for Kodikanal data.}
\label{table2}
\end{table}

Without the influence of convection, untwisted flux tubes tilt in the appropriate direction for their respective hemispheres due to the Coriolis force acting on the limbs of the emerging loop ({\it{e.g.}} \opencite{dsilva93}).  However, for weaker initial field strengths of 15 kG and 30 kG, some emerging loops show negative tilt angles, opposite to the sign of the active region mean tilts.  This occurs because plasma flow along the flux tube near the apex changes from diverging to converging as it enters the upper convection zone ({\it{e.g.}} \opencite{cali95}; \opencite{fan_fisher1996}).  The Coriolis force acting on the converging flow drives a tilt of the wrong sign, resulting in a negative best fit line slope for tubes without convection at magnetic field strengths of 15 kG and 30 kG.  This effect is especially severe here for flux tubes of 15 kG and $10^{20}$ Mx.  The tilt angle of these tubes is so large and of the wrong sign for initial latitudes of $5^{\circ}$, $10^{\circ}$, and $15^{\circ}$, that they do not appear on the tilt angle plot (Figure \ref{fig:tiltangle}-(a), lower left).  As can be seen from Table \ref{table1} and Table \ref{table2}, we find that at all magnetic fields, the slopes of the best-fit lines for the flux tubes with convection are increased from the cases without convection.  This occurs in part because the upflows in the convective velocity field have an associated kinetic helicity that helps to drive the tilt of the flux tube apex in the appropriate Joy's Law direction for its respective hemisphere (i.e. toward the equator).  Such a mean kinetic helicity corresponds to a vertical vorticity in upflows that is clockwise in the northern hemisphere, and counter-clockwise in the southern hemisphere.  

With convection and for all magnetic fluxes, all magnetic field strengths of 15 kG\,--\,100 kG show positive best-fit line slopes, agreeing with Joy's Law.  This is an improvement upon Article 1, for which there were fewer simulations only for a magnetic flux of $10^{22}$ Mx and the best-fit line slopes for 15 kG and 30 kG of 0.12 $\pm$ 0.18 and 0.15 $\pm$ 0.21 respectively, had too large uncertainties to report a definitive Joy's Law trend.  It is interesting to note that with convection, the mid-magnetic field strength flux tubes (40 kG\,--\,50 kG) have the largest best-fit slopes.  This is probably due to the fact that at these mid-field strengths, the joint effects of magnetic buoyancy and the convective flows are such that the tubes have a longer rise time compared to the other field strengths. Therefore, the systematic effects from the Coriolis force and kinetic helicity in convective upflows have a longer time to act on the flux tubes, and the flux tube will likely encounter multiple convective cells during its evolution.  Since the mean kinetic helicity is obtained by averaging the kinetic helicity of many cells over time, a flux tube which takes a longer time to emerge will be influenced more by the mean kinetic helicity rather than stochastic fluctuations.

\begin{figure}    
   \centerline{\hspace*{0.015\textwidth}
               \includegraphics[width=0.41\textwidth,clip=]{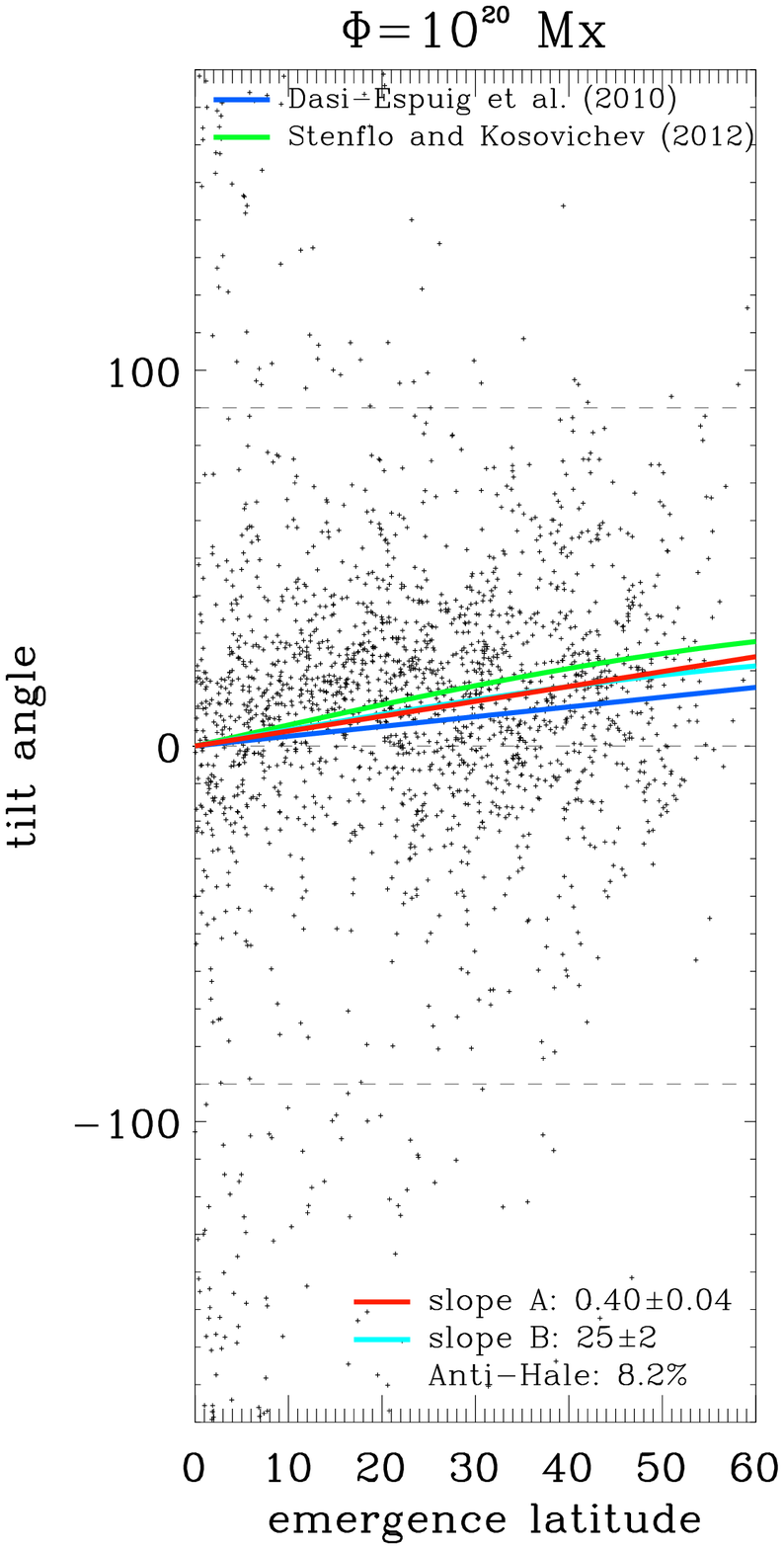}
               \hspace*{-0.03\textwidth}
               \includegraphics[width=0.41\textwidth,clip=]{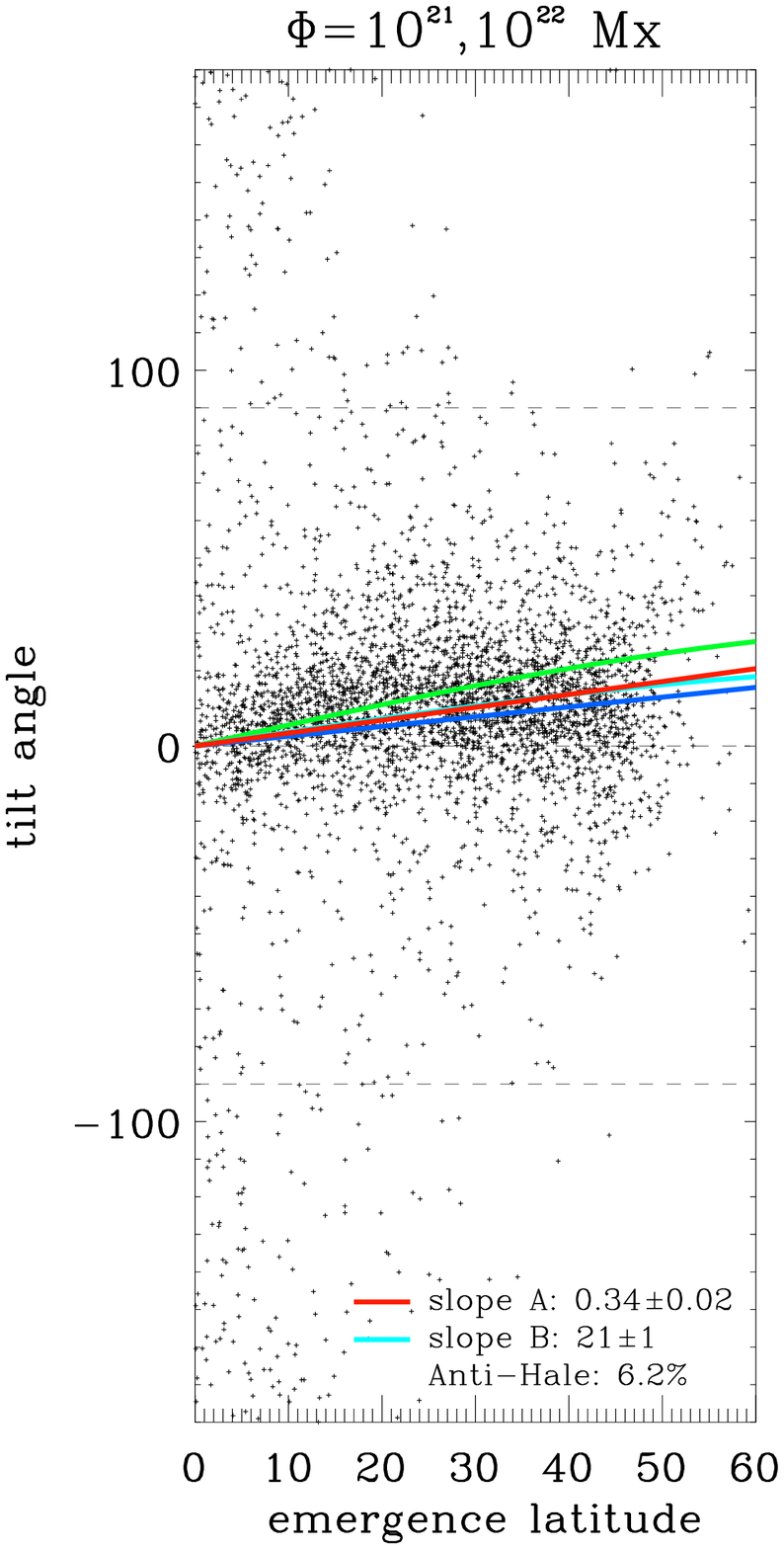}
              }
     \vspace{-0.75\textwidth}   
     \centerline{\large \bf     
      \hspace{0.15 \textwidth}  \color{black}{(a)}
      \hspace{0.32\textwidth}  \color{black}{(b)}
         \hfill}
     \vspace{0.75\textwidth}    
   \centerline{\hspace*{0.015\textwidth}
               \includegraphics[width=0.41\textwidth,clip=]{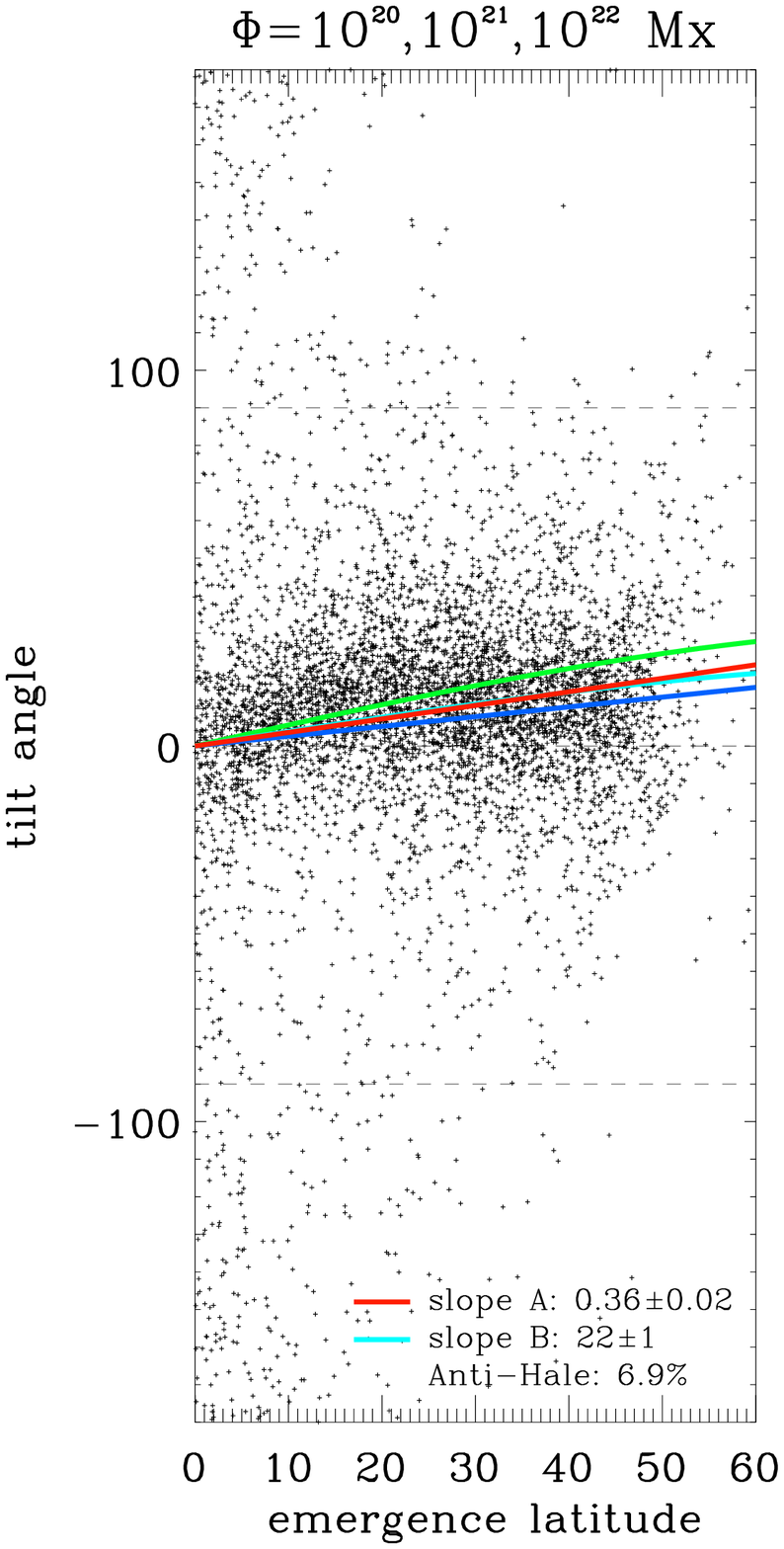}
               \hspace*{-0.03\textwidth}
               \includegraphics[width=0.41\textwidth,clip=]{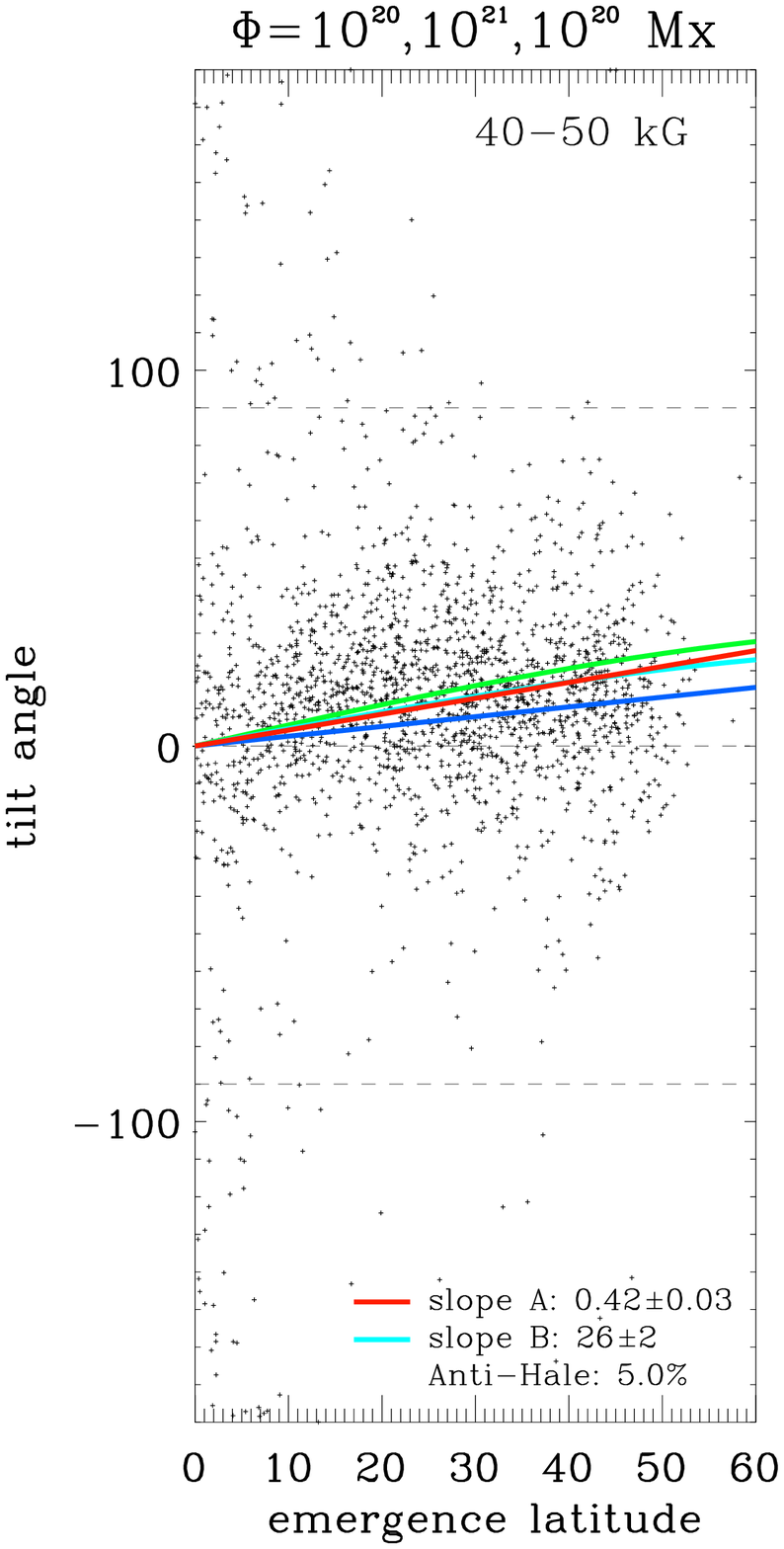}
              }
     \vspace{-0.75\textwidth}   
     \centerline{\large \bf     
      \hspace{0.15 \textwidth} \color{black}{(c)}
      \hspace{0.32\textwidth}  \color{black}{(d)}
         \hfill}
     \vspace{0.75\textwidth}    
              
\caption{Tilt angles for various flux and magnetic fields plotted together. The dark blue line is the linear best-fit line for tilt angle as a function of latitude as found from white light sunspot group observations (Dasi-Espuig $\it{et}$ $ \it{al.}$, 2010), and the green line is the best-fit line for tilt angle as a function of the sine of latitude as found from magnetograms (Stenflo and Kosovichev, 2012).  In all graphs, we also plot the best-fit line of the simulations for tilt angle as a function of latitude (red, slope A), and for the tilt angle as a function of the sine of latitude (light blue, slope B) for (a) tilt angles of all magnetic field strengths and flux of $10^{20}$ Mx only, (b) tilt angles for all magnetic field strengths and fluxes of $10^{21}$ Mx and $10^{22}$ Mx, (c) tilt angles for all magnetic field strengths and fluxes of $10^{20}$ Mx, $10^{21}$ Mx, and $10^{22}$ Mx, (d) tilt angles for all fluxes  for magnetic field strengths of 40 kG\,--\,50 kG.  The best-fit lines for the thin flux tube simulations are bounded by the observational best-fit lines.  We find that 6.9$\%$ of our simulated flux tubes are of an anti-Hale configuration, as compared to the $\approx$4$\%$ as found by Stenflo and Kosovichev (2012).}
   \label{fig:tiltall}
   \end{figure}

For solar observational data, the observed tilt angle of the active region is often plotted as a function of its emergence latitude for active regions of all sizes and magnetic flux.  Figure \ref{fig:tiltall} shows the tilt angles from the flux tube simulations together for selected values of magnetic field and flux.  A linear least-squares fit is performed on this data for various flux and magnetic field combinations as shown in Figure \ref{fig:tiltall}, using the fit of $\alpha=m_{A}\lambda$.  From white light sunspot group tilt angle data spanning solar cycles 15 -- 21, \inlinecite{espuig10} find a linear best-fit line slope $m_{A}$ of 0.26 $\pm$ 0.05 for Mount Wilson data, and 0.28 $\pm$ 0.06 for Kodikanal data.  For simulated flux tubes of all magnetic field strengths considered and fluxes of $10^{21}$ Mx and $10^{22}$ Mx (Figure \ref{fig:tiltall}-(b)), as these are most likely the magnetic flux strengths required to produce sunspot groups that can be identified in white light, we find a slope of $m_{A}=0.34$ $\pm$ $0.02$, which does fall within the range of tilt angles that \inlinecite{espuig10} find given the uncertainties on the slope.

A Joy's Law fit can also be performed using $\alpha=m_{B}$sin$(\lambda)$, which is a good choice assuming the origin of the tilt angle is related to the Coriolis force, as this force varies with latitude as sin$(\lambda)$.  Here, $m_{B}$ is the best-fit line slope. \inlinecite{stenflo12} perform such a fit, which gives a slope of $m_{B}=32.1^{\circ}$ $\pm$ $0.7^{\circ}$, using 15 years of MDI full-disk magnetograms.  \inlinecite{wang89} find a similar trend using National Solar Observatory Kitt Peak magnetograms of bipolar active regions for Solar Cycle 21.  Performing the same fit on the tilt angles of our simulated flux tubes for magnetic fluxs of $10^{20}$ Mx, $10^{21}$ Mx, and $10^{22}$ Mx (Figure \ref{fig:tiltall}-(c)), which encompasses the range of flux these magnetograms consider, we find $m_{B}=22^{\circ}$ $\pm$ $1^{\circ}$.  This value is much lower than what \inlinecite{wang89} and \inlinecite{stenflo12} determine.  However, using Mount Wilson white light sunspot group data collected from 1917 -- 1985, \inlinecite{fisher95} find $m_{B}=15.69^{\circ}$ $\pm$ $0.66^{\circ}$.  For our simulated flux tubes of $10^{21}$ Mx and $10^{22}$ Mx (Figure \ref{fig:tiltall}-(b)), we find $m_{B}=21^{\circ}$ $\pm$ $1^{\circ}$, which is slightly larger than the value derived by \inlinecite{fisher95}.  Recalling that mid-field strengths of  40 kG -- 50 kG have the largest linear best-fit slopes $m_{A}$, we perform the fit $\alpha=m_{B}$sin$(\lambda)$ only for these field strengths and all magnetic fluxes we consider (Figure \ref{fig:tiltall}-(d)).  The result is a significantly increased value of $m_{B}=26^{\circ}$ $\pm$ $2^{\circ}$, although it still falls short of the $32.1^{\circ} \pm 0.7^{\circ}$ value found by \inlinecite{stenflo12}.  For all values of magnetic flux and magnetic field that we consider for Figures \ref{fig:tiltall}-(a) through \ref{fig:tiltall}-(d), the best-fit lines we obtain for our simulated flux tubes are bounded by the tilt angle best-fit lines as derived from observations.  We note that the best-fit lines for the Joy's Law tilt angle dependence derived from white light sunspot group data (\opencite{fisher95}; \opencite{espuig10}) are much lower than those derived from magnetograms (\opencite{wang89}; \opencite{stenflo12}).  This might occur because weaker active regions can be identified in magnetograms, which may not appear in white light sunspot group images.  However, it may also be the result of selection effects employed by various authors, as well as the fact that anti-Hale regions might not be identified from white light data.

According to \inlinecite{fan94}, in the absence of convection, the tilt angle of the flux tube should increase with increasing flux as a result of the Coriolis force acting on the buoyantly rising flux tube.  With convection, given the uncertainties that we obtain, we do not find a statistically significant dependence of the Joy's Law slope on magnetic flux (see Table \ref{table2}), which agrees with the results of \inlinecite{stenflo12}.  We have shown that the addition of convection increases the Joy's Law slope, regardless of the amount of time that the flux tubes stays in the bulk of the convection zone.  This indicates that convective effects can have a significant contribution to the tilt angle of the flux tube throughout its evolution.  Therefore, the results of \inlinecite{stenflo12} do not rule out the paradigm that flux tubes obtain at least a portion of their tilt angle during their buoyant rise.

We also note that 6.9$\%$ of our simulated flux tubes (see Figure \ref{fig:tiltall}-(c)) emerge with polarites that violate Hale's Law, in comparison to the $\approx$$4\%$ as found via observations of medium to large-sized active regions (\opencite{wang89}; \opencite{stenflo12}).  In our simulations, these violations arise as a result of flux tubes emerging in the opposite hemisphere from which they originated, or as a result of the flux tube becoming so distorted by convection that the legs of the emerging loop can become reversed.  Both of these situations usually occur for weak magnetic field strength cases.  The majority of emerging anti-Hale flux tubes with moderate initial magnetic fields between 40 kG\,--\,50 kG happen as a result of hemisphere crossing due to convective flows.  Therefore, in light of our findings, observations of anti-Hale spots of various magnetic flux does not rule out the rising flux tube paradigm.

\subsection{Tilt Angle Scatter}
We attempt to further constrain the magnetic field strength at which the solar dynamo might be operating by observing the scatter of tilt angles our simulated flux tubes produce about the Joy's Law trend.  In Figure \ref{fig:tiltangle}, there is a clear scatter of the tilt angle about the best-fit line for the data with convection, which becomes greater at lower magnetic flux.  Again, this trend is the result of stronger coupling between the flux tube and convective flows at a reduced magnetic flux.  To quantify the scatter of the tilt angles around the best fit line, we calculate the standard deviation of the tilt about its fitted value:

\begin{equation}
\sigma=\sqrt{\frac{ \sum_{i=1}^{N} (\alpha_{i}-\alpha_{fit})^{2}}{N}},
\end{equation} 
where $\alpha_{i}$ is the $i^{th}$ tilt angle, $\alpha_{fit}$ is the $i^{th}$ tilt angle as a result of the fit for which the slopes $m_A$ are reported in Table \ref{table2}, and $N$ is the number of points considered.

Using Mount Wilson white light sunspot group data (1917\,--\,1985),  \linebreak \inlinecite{fisher95} find that $\sigma < 40^{\circ}$ for all emergence latitudes. We evaluate $\sigma$ for each field strength and flux value, and the results are given in Table \ref{tbl:scatter}.  Comparing to the results of  \inlinecite{fisher95}, this would exclude magnetic field strengths of 15 kG and 30 kG for the combination of fluxes of $10^{21}$ Mx and $10^{22}$ Mx together (row 4 in Table \ref{tbl:scatter}) as the progenitors of solar active regions, which has a scatter that exceeds the observed value.  We only consider these flux values because \inlinecite{fisher95} uses white light sunspot group data.  If we include all magnetic fields and these two fluxes together (not shown in Table \ref{tbl:scatter}), we obtain a $\sigma$ value of 41.4 $\pm$ 0.7, which is fairly close to what \inlinecite{fisher95} obtain.  However, if we exclude magnetic fields of 15 kG and 30 kG for fluxes of $10^{21}$ Mx and $10^{22}$ Mx (not shown in Table \ref{tbl:scatter}), we obtain a $\sigma$ value of 25.3 $\pm$ 0.5, well within the range that \inlinecite{fisher95} suggests.  

Active regions of flux values of $10^{20}$ Mx would appear on magnetograms, but would probably not be associated with white light sunspot groups.  Since we have simulations for $10^{20}$ Mx, we calculate $\sigma$ for these values as well in Table \ref{tbl:scatter}.  For this magnetic flux, $\sigma$ is very large and exceeds the values suggested by \inlinecite{fisher95} for all magnetic field strengths $\le$40 kG.  However, this result does not exclude these magnetic field values as progenitors of active regions for a flux of $10^{20}$ Mx, since these regions most likely would not appear in the \inlinecite{fisher95} study.  We do note that while the Joy's Law trend is not statistically dependent on flux (see Table \ref{table2}), the scatter of the tilt angles about their best-fit line tends to increase with decreasing flux (see Table \ref{tbl:scatter}).  \inlinecite{wang89} and \inlinecite{stenflo12} also find that the spread in the distribution of the tilt angles increases as one goes to smaller regions, or smaller magnetic flux.

\begin{table}
\begin{tabular}{ccccccccc}

\hline
                           & 15 kG                   & 30 kG                    & 40 kG                & 50 kG  
                           & 60 kG                   &100 kG\\
\hline
$10^{20}$ Mx  & 63 $\pm$ 4      & 58 $\pm$ 3      & 54 $\pm$ 3     & 41 $\pm$ 2
                           & 40 $\pm$ 2      & 22 $\pm$ 1\\

$10^{21}$ Mx  & 73 $\pm$ 4      & 55 $\pm$ 3      & 42 $\pm$ 2     & 30 $\pm$ 2
                            & 19 $\pm$ 1    & 10.4 $\pm$ 0.6\\

$10^{22}$ Mx  & 71 $\pm$ 4      & 52 $\pm$ 3      & 35 $\pm$ 2     & 23 $\pm$ 1
                            & 12.5 $\pm$ 0.7     & 8.2 $\pm$ 0.4\\

$10^{21}$, $10^{22}$ Mx  & 74 $\pm$ 2      & 54 $\pm$ 2      & 39 $\pm$ 2     & 26 $\pm$ 1
                            & 16.3 $\pm$ 0.6     & 9.4 $\pm$ 0.4\\

$10^{20}$, $10^{21}$, $10^{22}$ Mx  & 69 $\pm$ 2      & 55 $\pm$ 2      & 44 $\pm$ 1     & 32 $\pm$ 1
                            & 26.4 $\pm$ 0.8     & 14.9 $\pm$ 0.5\\
\hline
\end{tabular}
\caption{Standard deviation of the tilt angle about its fitted value.  The standard deviation tends to increase with decreasing flux.}
\label{tbl:scatter}
\end{table}

\subsection{Preferred Tilt Angle}
\begin{figure}
\centerline{\includegraphics[width=0.7\textwidth,clip=]{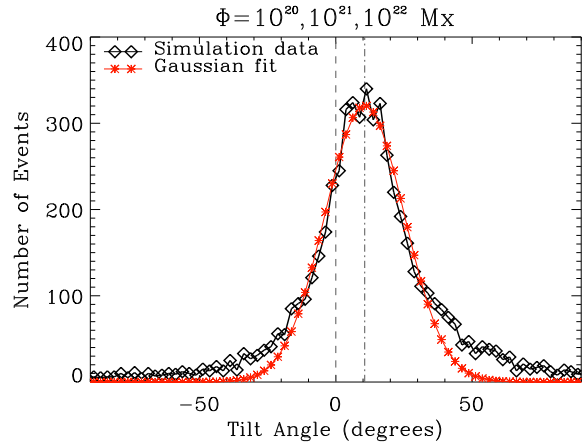}}
\caption{Distribution of tilt angles in 2.5$^{\circ}$ bins for all initial magnetic field strengths and all magnetic fluxes considered in this study.  The peak of the distribution function is between 10$^{\circ}$ -- 12.5$^{\circ}$ (black line), with a non-linear least-squares Gaussian fit (red line) of center 10.6$^{\circ}$, as shown by the dash-dotted line.}
\label{fig:histo}
\end{figure}

\inlinecite{howard96} investigated the distribution of sunspot group tilt angles from a set of Mount Wilson white light photographs (1917\,--\,1985) as well as sunspot groups derived from plage in Mount Wilson daily magnetograms (1967\,--\,1995).  They bin the tilt angles in 2.5$^{\circ}$ increments, and plot the number of ocurrances of the tilt angles in these bins.  In this way, they find that sunspot groups tend to form most often with tilt angles between 2.5$^{\circ}$\,--\,5$^{\circ}$ for white light data, and between 7.5$^{\circ}$\,--\,10$^{\circ}$ for magnetogram data.  The tilt angle distribution that we obtain for flux tubes of all magnetic fields and magnetic flux studied here using tilt angle bins of 2.5$^{\circ}$ is shown in Figure \ref{fig:histo}. We find that for all magnetic fluxes considered, the tilt angle distribution peaks between bins of 10$^{\circ}$\,--\,12.5$^{\circ}$ (black line), where a non-linear least-squares Gaussian fit (red line) gives a center, or preferred tilt, of 10.6$^{\circ}$. This is in good agreement with the magnetogram data of \inlinecite{howard96} and \inlinecite{stenflo12}, who find a preferred, or most common, tilt of $\approx$10$^{\circ}$ for all bipolar regions emerging between a 15$^{\circ}$\,--\,20$^{\circ}$ latitude range including the largest active regions down to regions of $\approx10^{20}$ Mx.

\inlinecite{howard96} also finds an average tilt angle of 4.3$^{\circ}$ for sunspot groups and 6.3$^{\circ}$ for the magnetograms, whereas we find an average of $9.5^{\circ}\pm0.6^{\circ}$ for all fluxes considered together, where the uncertainty is the standard deviation of the mean.  The average tilt angle value is significantly larger than what \inlinecite{howard96} finds.  However, this could be due to the nature of our chosen convective velocity simulations, the thin flux tube idealization, or a result of the selection criteria that \inlinecite{howard96} uses to identify sunspot groups or plages.  They count an active region on every day it is observed, so an active region may be represented a number of times for a single solar disk passage.  As such, their data also includes the evolution of the region's tilt angle as it ages.       

\subsection{Magnetic Fields at the Flux Tube Apex}
Using thin flux tube simulations, \inlinecite{fan93} show that the preceeding leg of an emerging flux loop has a stronger magnetic field than the following leg as a result of the Coriolis force, which induces a differential stretching of the rising loop.  This provides an explanation for the observed coherent, less fragmented morphology for the leading polarity flux of an active region ({\it{e.g.}} \opencite{bray_loughhead1979}). Subsequent simulations using the more physical mechanical equilibrium state ({\it{e.g.}} \opencite{cali95}, \opencite{fan_fisher1996}), as opposed to temperature equilibrium (\opencite{fan93}), found that the leading leg of the emerging loop has a stronger magnetic field than the following only for flux tubes with an initial field strength below 60 kG.  As is in Article 1, we investigate the magnetic field asymmetry by calculating $dB/ds$, the derivative of the magnetic field along the arc length $s$ in the direction of solar rotation, at the apex of the emerging loop. If $dB/ds$ is greater (less) than zero, then the leading (following) leg has a stronger magnetic field.  We find that for all fluxes, tubes with an initial magnetic field $\leq 50$ kG tend to emerge with the appropriate magnetic field asymmetry of a stronger field strength in the leading side, consistent with the tendency of a more coherent leading polarity in an active region.

Recent three-dimensional radiation MHD simulations of magnetic flux emergence and sunspot formation in the top layer of the convection zone and the photosphere by Rempel (private communication, 2012) have suggested that the morphological asymmetry of sunspots may be caused by a flow of plasma along the emerging tube out of the leading portion of the active region into the following leg, and is less dependent on the asymmetry in field strength of the emerging loop.  If this is indeed the case, then we can not rule out flux tubes with stronger initial magnetic field strengths  of $> 50$ kG at the base of the solar convection zone as the progenitors of solar active regions, simply due to their magnetic field asymmetries.     

\begin{figure}
\centerline{\includegraphics[width=0.95\textwidth,clip=]{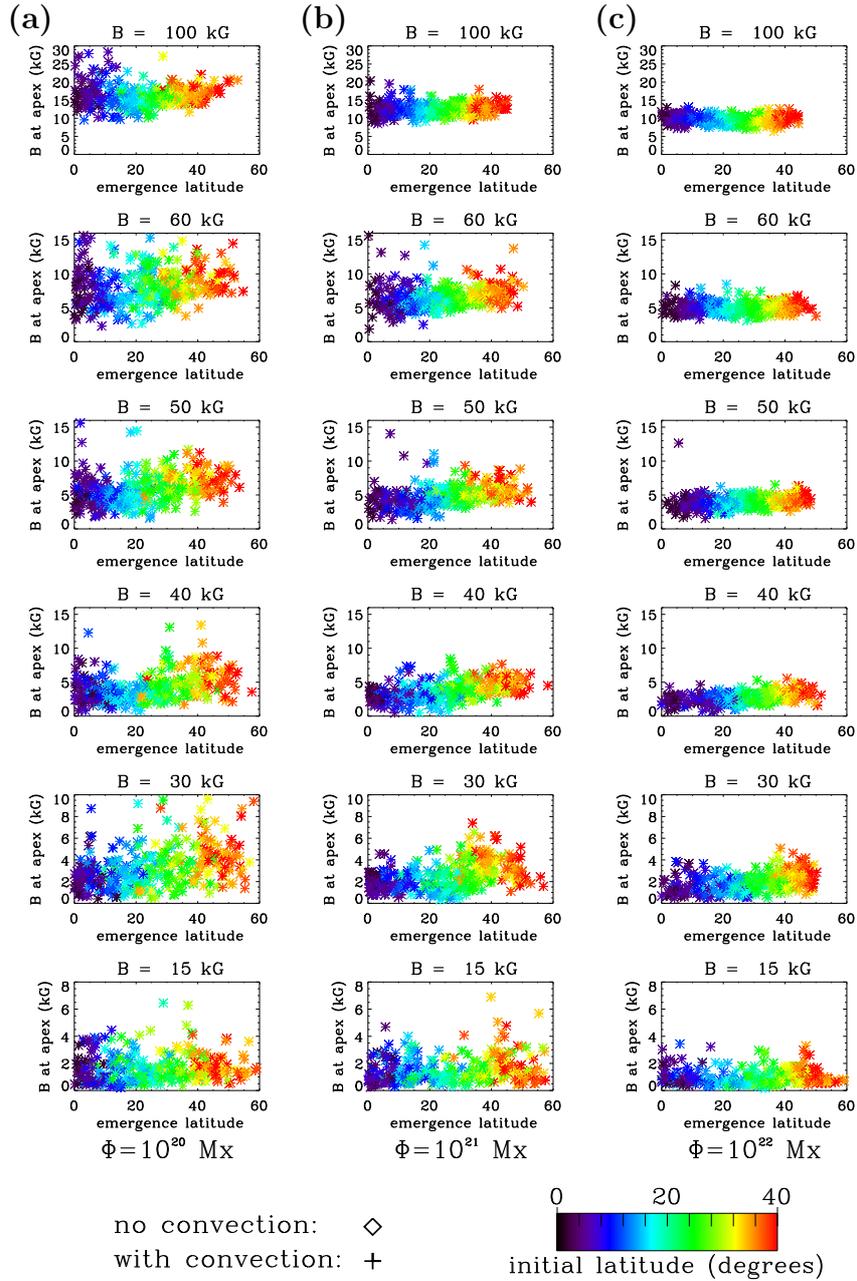}}
              \vspace{-1.45\textwidth}   
     \centerline{\large \bf     
      \hspace{0.03 \textwidth}  \color{black}{(a)}
      \hspace{0.24\textwidth}  \color{black}{(b)}
      \hspace{0.24\textwidth}  \color{black}{(c)}
       \hfill}
     \vspace{1.45\textwidth}    

\caption{Magnetic field strength at the apex of the flux loop for (a) $10^{20}$ Mx in column 1, (b) $10^{21}$ Mx in column 2, (c) and $10^{22}$ Mx in column 3.  Our results support average magnetic field strengths of $\approx$500 G to $\approx$15 kG for flux tubes apices which have reached the top of the simulation domain, which is still 25 Mm below the solar surface.}
\label{fig:bee}
\end{figure}

The nature of the magnetic field below the visible surface of the Sun is not well known.  Simulations of flux emergence in the upper layers of the convection zone have been computed in domains that span from the photosphere to depths of 2 Mm to 20 Mm below the surface ({\it{e.g.}} \opencite{cheung07}; \opencite{rempel09}; \opencite{stein11}).  The simulations assume a range of initial magnetic field strengths of the flux tube at the bottom of their simulation domain anywhere from $\approx$2 kG to $\approx$30 kG.  We show the magnetic field strength at the apex of our thin flux tube simulations, once the tube has reached the top of the simulation domain at a depth of 25 Mm below the solar surface, for multiple initial magnetic field strengths and flux in Figure \ref{fig:bee}.  Our results do support average magnetic field strength values of $\approx500$ G  to $\approx15$ kG for flux tubes reaching a height of $\approx$21 Mm below the solar surface, assuming initial field strengths of 15 kG to 100 kG at the base of the convection zone for magnetic fluxes of $10^{20}$ Mx to $10^{22}$ Mx.  The magnetic field strength of the flux tube at the apex is largest for larger initial fields, and the scatter of the magnetic field at the apex for a particular magnetic field strength increases with decreasing flux.

\subsection{Emerging Loop Rotation Rate}
\label{sec:rotrate}
\begin{figure}
\centerline{\includegraphics[width=0.6\textwidth,clip=]{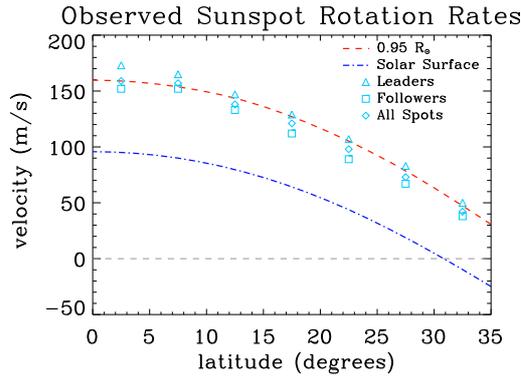}}
\caption{Rotation rate of leader, follower, and all sunspots, plotted with the observed azimuthal rotation rate of the solar plasma at the surface (blue dash-dotted line) and $r = 0.95 R_{\odot}$ (red dashed line).  All values are plotted with reference to the solid body rotation rate of $\Omega_{0}$=$2.7 \times 10^{-6}$ rad s$^{-1}$, so that the zero line is the solid body rotation of the Sun.  This image suggests that sunspots rotate at nearly the same rate as the solar plasma at $0.95 R_{\odot}$.}
\label{fig:spots}
\end{figure}

Observations show that sunspots tend to rotate faster than the solar surface plasma \cite{howard1970,golub1978}.   The azimuthal \linebreak velocity of leader, follower, and all sunspots (in the reference frame with a solid body rotation rate of $\Omega_0 = 2.7 \times 10^{-6}$ rad s$^{-1}$) as derived from \linebreak \inlinecite{gilman85} are plotted in Figure \ref{fig:spots}, as well as the observed azimuthal rotation rate of the solar plasma at the surface (blue line)  as determined from surface spectroscopic Doppler-velocity measurements \linebreak ({\it{e.g.}} \opencite{thompson2003}) and the rotation rate at $r=0.95 R_{\odot}$ (red line) as found via inversions of helioseismic observations \cite{howe00}. This image suggests that sunspots tend to rotate at nearly the same rate as the solar plasma at $0.95 R_{\odot}$, and therefore rotate faster than solar surface plasma.  This might also indicate that sunspots are anchored at a depth of $0.95 R_{\odot}$, near the surface shear layer interface. 

In order to quantify the rotation rate of emerging loops from our thin flux tube simulation, we calculate the apparent azimuthal speed of the center point between the leading and following intersections of the emerging loop with the constant $r$ surface of $0.95 R_{\odot}$ during the last two time steps before the loop apex reaches the top of the simulation domain. This reflects the apparent azimuthal speed of an emerging active region, and it is not the actual azimuthal speed of the flux tube plasma at the tube apex.  We do this only for flux tubes of $10^{21}$ Mx and $10^{22}$ Mx, assuming tubes of flux order $10^{20}$ Mx do not produce large sunspots. \inlinecite{cali95} computed the azimuthal phase speed of the summit of an emerging flux loop without convective effects throughout its rise.  This phase motion is related to the wave character of the rising flux loop, which is similar in behavior to a transversal wave propagating along a string.  They find that the azimuthal phase speed decreases with increasing height of the summit, and that it changes sign at about 50 Mm below the solar surface such that the angular velocity of the summit is smaller than that of the external plasma.  However, they suggest that the resulting active region will still show a higher rotation velocity than the surrounding plasma because of the inclination difference of the leading and following legs with respect to the local vertical.  This will occur when the following leg has a steeper slope than the leading leg, which is caused by the conservation of angular momentum as the tube rises (\opencite{mi94}; \opencite{cali95}; \opencite{cali98}).  In Article 1, we find that, on average, all magnetic field strengths between 15 kG and 100 kG at a magnetic flux of $10^{22}$ Mx show the appropriate inclination asymmetry of the emerging flux loop such that the leading and following polarities of a resulting active region will appear to drift apart from each other and rotate faster than the surrounding plasma.  To make comparisons to actual sunspot rotation rates, we feel that the method presented in this section is a better alternative.

\begin{figure}
 \centerline{\hspace*{0.015\textwidth}
               \includegraphics[width=0.515\textwidth,clip=]{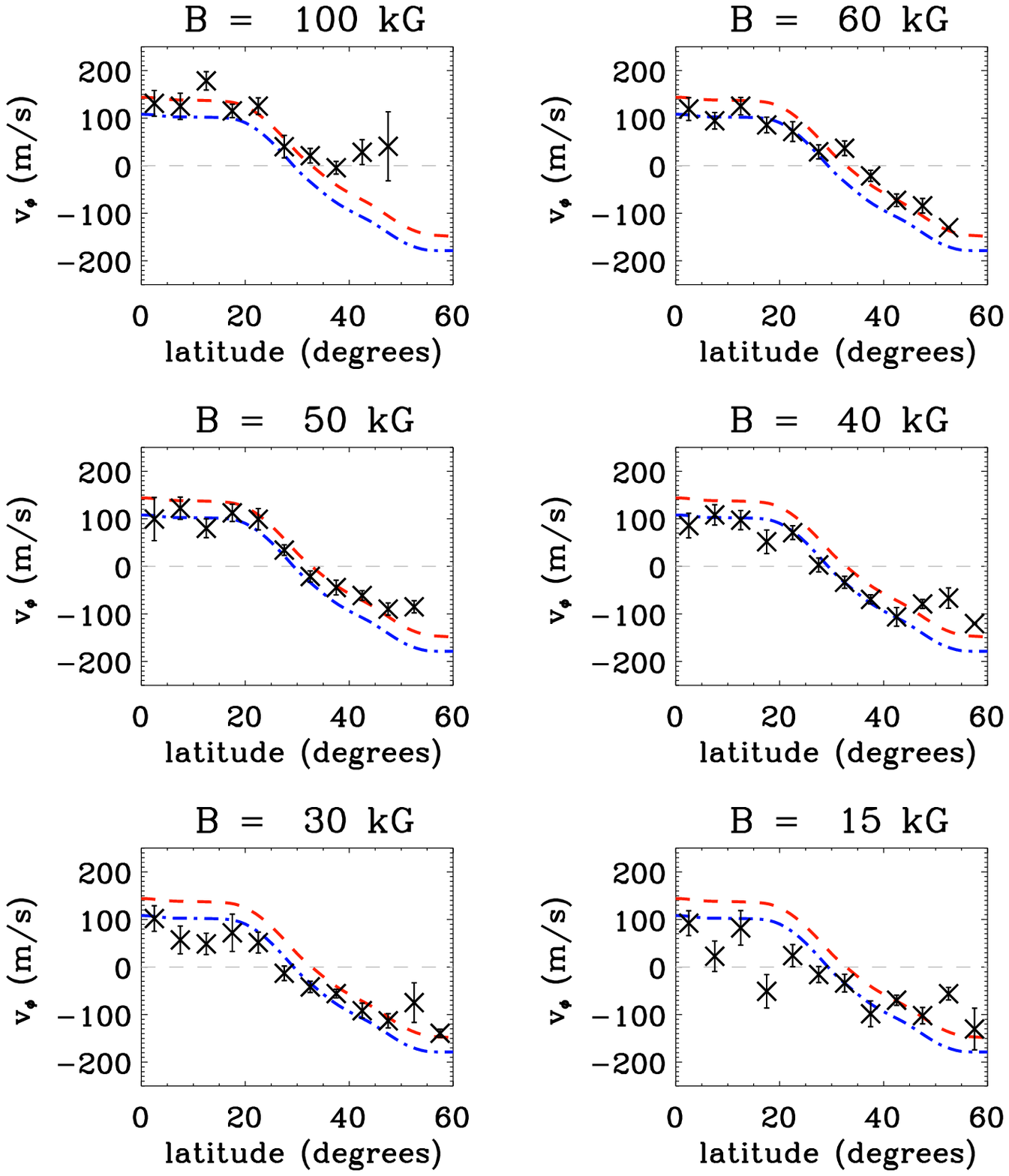}
               \hspace*{-0.03\textwidth}
               \includegraphics[width=0.515\textwidth,clip=]{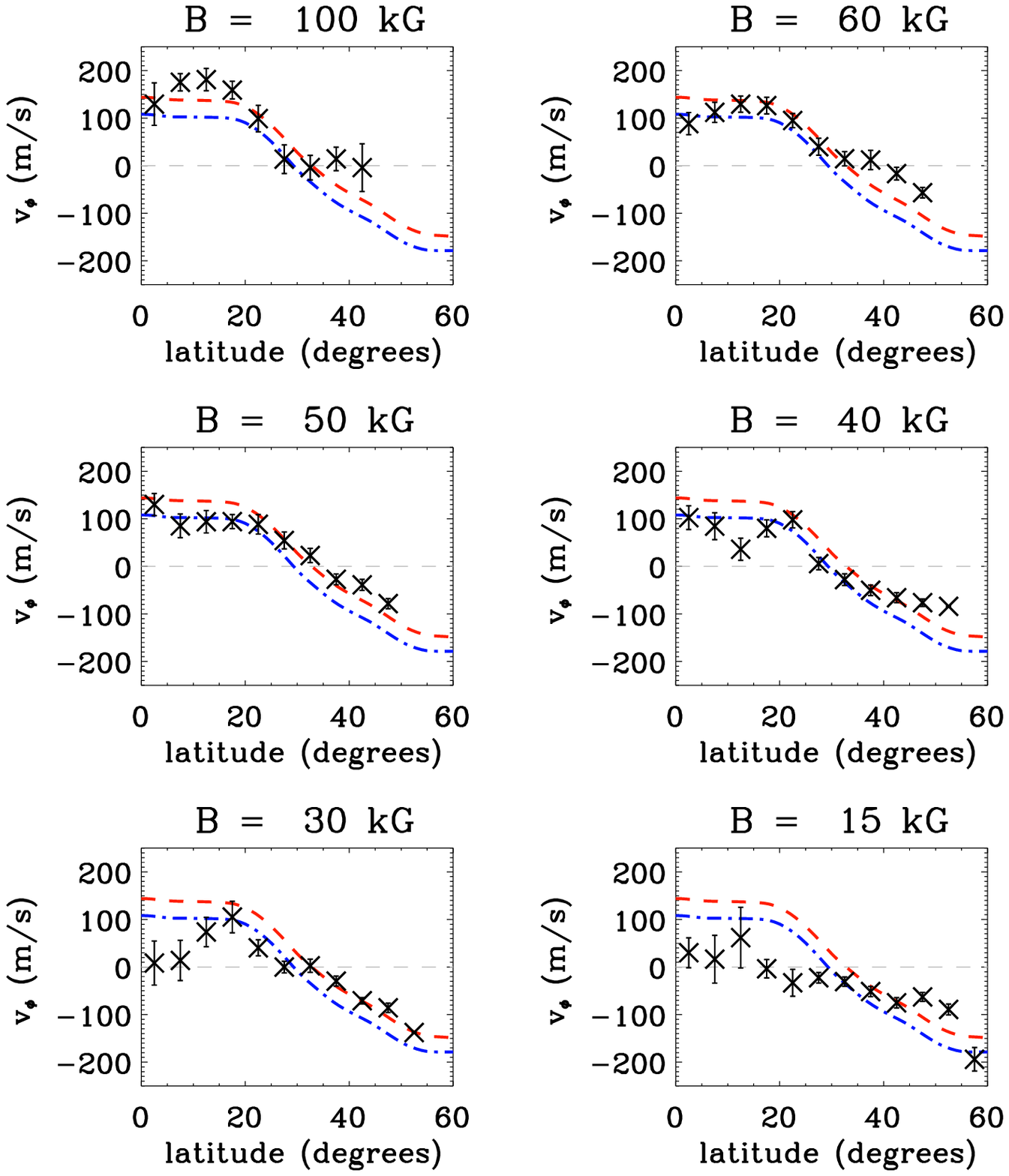}
              }
                \vspace{-0.615\textwidth}   
     \centerline{\small \bf     
      \hspace{0.0 \textwidth}  \color{black}{(a)}
      \hspace{0.445\textwidth}  \color{black}{(b)}
         \hfill}
     \vspace{0.6\textwidth}    
   \caption{Average rotation rate of emerging flux loops subjected to the convective flow with magnetic flux of (a) $10^{21}$ Mx on the right (rightmost six panels) and (b) $10^{22}$ Mx on the left (leftmost six panels).  Averages are taken in 5$^{\circ}$ bins, with bars representing the standard deviation of the mean.  Red and blue lines correspond to the ASH rotation rate at $r = 0.95 R_{\odot}$ and inferred solar surface rotation rate respectively, assuming a difference of 10 nHz between the two. All values are plotted with reference to the solid body rotation rate of $\Omega_{0}$ = $2.7 \times 10^{-6}$ rad s$^{-1}$, so that the zero line is the solid body rotation of the Sun.  Strong flux tubes with initial field strengths of $\ge$ 60 kG rotate faster than or nearly equal to the inferred surface rate.}           
\label{fig:rotrate}
\end{figure}

Figures \ref{fig:rotrate}-(a) and \ref{fig:rotrate}-(b) show the average of the apparent azimuthal speed in $5^{\circ}$ bins, for cases with a magnetic flux of $10^{21}$ Mx and $10^{22}$ Mx respectively, where bars on the points are the standard deviation of the mean.  For comparison, these plots also show the average azimuthal rotation rate of the ASH simulation at $r = 0.95 R_{\odot}$ (red line), and the azimuthal flow speed one would expect at the surface (blue line) assuming the differential rotation in the top shear layer between the surface and $0.95 R_{\odot}$ decreases by $\approx$10 nHz (what we will subsequently call the \emph{inferred} surface rate) as found by helioseismology ({\it{e.g.}} \opencite{thompson2003}).         

In all cases, flux tubes that emerge at high latitudes above 40$^{\circ}$ rotate at a rate that is close to, or faster than, the ASH rate at $r=0.95 R_{\odot}$ and hence faster than the inferred surface rate expected of our simulation.  With initial magnetic field strengths of 15 kG and 30 kG, for both magnetic fluxes, the average rotation rates of the majority of flux tubes that emerge at $\le$ 35$^{\circ}$ are less than the inferred surface rate.  However, at mid-magnetic-field strengths of 40 kG and 50 kG, the rotation rates of the flux tubes roughly follow the inferred surface rate.  Only for initial magnetic field strengths of $\ge$ 60 kG are the flux tubes capable of rotating at or faster than the inferred surface rate for all emergence latitudes, considering magnetic fluxes of $10^{21}$ Mx and $10^{22}$ Mx.

We recognize that this method of quantifying the rotation rate of the emerging flux tubes in our simulation has some limitations.  Due to the nature of the ASH convection simulation and the fact that the thin flux tube simulation breaks down in the upper portion of the convection zone, it is not possible to allow the thin flux tube to emerge all the way to the solar surface.  Instead, we operate under the assumption that the rotation rate of the emerging loop at a constant {\it{r}} surface of $0.05 R_{\odot}$ below the solar surface is a good representation for how an active region will behave at the solar surface, as is reflected in Figure \ref{fig:spots} for solar observations.  The discrepancy between the observed and simulated rotation rates could in part be attributed to the particular ASH simulation that we use, which does not precisely reproduce the solar $\Omega$-profile.  Our simulated flux tubes are effectively anchored at the convection zone base, whereas real flux tubes may decouple from the deep convection zone at some point during their evolution, and become anchored closer to the surface.  This is why we choose to investigate the rotation rate closer to the surface.  As our thin flux tube is a one-dimensional string of mass elements, we cannot address the issue that the flux tube could lose its coherency and become fragmented (\opencite{longcope96}), which may result in a stronger coupling between the tube and convective fluid motions, and could be a significant contributing factor to the rotation rate of active regions.

In an attempt to gain a stronger coupling between the flux tube and convective fluid motions, we perform some simulations where the drag coefficient $C_{d}$ in the last term of Equation (\ref{eq:eqn_motion}) is increased from unity to the constant values of 1.5 and 2.  Also, we perform simulations where we adjust the drag coefficient such that it exponentially varies from 1 at the base of the convection zone to 1.5 or 2 at the top of the simulation domain.  These tests were performed for flux tubes with initial field strengths of 40 kG and magnetic flux of $10^{22}$ Mx.  These efforts to alter the drag coefficient to produce stronger coupling to convection did not result in faster rotation rates compared to the surface rate for all latitudes less than 25$^{\circ}$.  This indicates that 40 kG flux tubes or less will probably not be able to reproduce sunspot rotation rates utilizing the thin flux tube model as it currently stands.  

\section{Summary}
\label{sec:discuss}

By embedding the thin flux tube model in a three-dimensional, turbulent, convective velocity field representing the solar convective envelope, we study how convection can influence the properties of emerging active region flux tubes.  In comparing these properties to those obtained from solar active region observations, we attempt to constrain the magnetic field strengths of dynamo-generated magnetic fields at the base of the solar convection zone. The thin flux tube approximation, although idealized, allows us to investigate active region scale flux tubes at weak to strong magnetic field strengths of 15 kG\,--\,100 kG under perfect frozen-in flux conditions.  We find that subjecting the thin flux tube to turbulent convective flows does indeed alter flux tube dynamics, and that it can have a significant impact on the properties of the emerging flux loop in comparison to flux tube simulations performed in the absence of a convective velocity field.  Also, the addition of convection aids the flux tube in more closely reproducing some observed properties of solar active regions.  This article extends upon the previous work of Article 1 by including flux tube simulations with additional magnetic fluxes of $10^{20}$ Mx, $10^{21}$ Mx, as well as $10^{22}$ Mx, and increasing the number of simulations performed per magnetic field strength by a factor of 3.5 in order to improve statistics such that uncertainties are reduced.  We also include more observational diagnostics, such as tilt angle scatter and sunspot rotation rate, to put further constrains on the field strength of the dynamo-generated magnetic field as the progenitors of solar active regions.

Decreasing the magnetic flux of the tube results in an increase of the drag force acting on the rising flux loop.  With convection, flux tube rise times decrease with decreasing flux for initial magnetic field strengths of 15 kG \,--\,60 kG because the increased drag force causes the flux tube to be more closely coupled with convection, and so they are advected by turbulent flows more strongly than the $10^{22}$ Mx case.  For all magnetic fluxes that we consider here, flux tubes are able to emerge near the equator with the aid of convective flows, which solves the previous problem of poleward slip for flux tubes of low magnetic field strengths without convective effects.

With the increased number of thin flux tube simulations, and hence improved statistics in this study, we are able to confidently say that for all magnetic field strengths of 15 kG to 100 kG, and all magnetic fluxes studied here, produce emerging loops with tilt angles that follow the Joy's Law trend.  This is an improvement upon Article 1 where the slope of the linear Joy's Law fit of the tilt angles of the emerging loops as a function of emerging latitude had too large an uncertainty for magnetic field strengths of 15 kG and 30 kG to report a definitive Joy's Law trend.  Of particular note is the fact that helical convective upflows help to drive the tilt angle of the flux tube in the appropriate Joy's law direction for both hemispheres. Including all tilt angles together for all magnetic field strengths and magnetic flux of $10^{21}$ Mx and $10^{22}$ Mx, we calculate the linear best-fit line for the tilt angle as a function of emergence latitude, and find a slope of $m_{A}=0.34$ $\pm$ $0.02$, which overlaps with the range of $0.26$ $\pm$ $0.05$ and $0.28$ $\pm$ $0.06$ as suggested by white light sunspot group image from Mount Wilson and Kodikanal, respectively (\opencite{espuig10}). Performing a fit for the tilt angle as a function of sine of the latitude for $10^{20}$ Mx, $10^{21}$ Mx, and $10^{22}$ Mx, we find a best-fit line slope of $m_{B}=22^{\circ}$ $\pm$ $1^{\circ}$, which is greater than the value $15.69^{\circ}$ $\pm$ $0.66^{\circ}$  obtained from white light sunspot group data by \inlinecite{fisher95}, but less than the value of $32.1^{\circ}$ $\pm$ $0.7^{\circ}$ found by \inlinecite{stenflo12} using MDI magnetograms.  When we exclude all fields except for 40 kG -- 50 kG for all the fluxes we consider here, $m_{B}$ increases significantly to $26^{\circ}$ $\pm$ $2^{\circ}$, which is closer to the value derived from magnetograms.

On the other hand, the scatter of the tilt angles around their linear Joy's Law fit line is shown to be too large for initial magnetic field strengths of 15 kG and 30 kG with fluxes of $10^{21}$ Mx and $10^{22}$ Mx, as compared to the observed value of \inlinecite{fisher95} for white light  sunspot group images.  While the scatter of the tilt angle increases with decreasing flux (Table \ref{tbl:scatter}), we find no statistically significant dependence of the Joy's Law trend on flux (Table \ref{table2}), consistent with the results of \inlinecite{stenflo12}.   We also find that the most common tilt angle produced by our study is 10.6$^{\circ}$ for tubes with a flux of $10^{20}$ Mx, $10^{21}$ Mx, and $10^{22}$ Mx, which agrees with \inlinecite{howard96} who finds that most tilt angles fall within the range of $7.5^{\circ}$\,--\,$10^{\circ}$ as obtained from Mount Wilson magnetogram data, although our average tilt angles are higher.

Similar to previous studies (\opencite{cali95}; \opencite{fan_fisher1996}; \linebreak \opencite{weber2011}), we find that for magnetic field strengths $\le$ 50 kG, the leading leg of the emerging loop tends to have a larger magnetic field than the following, which may provide an explanation for the observed better cohesion of the leading polarity of an emerging active region as compared to the following polarity.  This trend of asymmetry in field strength reverses for tubes with an initial magnetic field of $\ge$ 60 kG. However, it may be the case that the morphological asymmetry of sunspot regions is less dependent on magnetic field asymmetry, and is rather a result of the retrograde plasma flow inside the flux tube from the leading leg into the following leg as suggested by recent simulations of sunspot formation by Rempel (private communication, 2012).  If this is indeed the case, then we can not exclude magnetic field strengths of greater than 60 kG from the dynamo-generated magnetic field regime.  A study of the magnetic field of the flux tube at the top of the simulation domain suggests typical values of 500 G to 15 kG for tubes that reach $\approx$$21$ Mm below the photosphere.

Observations show that sunspot groups tend to rotate faster than the surrounding solar surface plasma (\opencite{howard1970}; \linebreak \opencite{golub1978}).  We use the apparent azimuthal motion of the center of the intersections of the emerging loop with a constant $r$ surface near the top of the simulation domain as a measure of the rotation rate of the emerging region and compare them to the average azimuthal rotation rate of the ASH convection simulation at $r=0.95R_{\odot}$, and the surface rotation rate that we would expect assuming the surface shear layer as inferred from helioseismology.  For tubes with a flux of $10^{21}$ Mx and $10^{22}$ Mx, we find that at high emergence latitudes, the average rotation rate of the emerging loops tends to be greater than the inferred surface rotation rate for all field strengths considered. At lower latitudes, below about $35^{\circ}$, loops with initial field strength $\ge$60 kG tend to rotate faster than the inferred surface rate, consistent with the observed sunspot rotation rate, while loops with initial fields of about 40 kG -- 50 kG tend to rotate at a similar rate as the surface rate.  However, for initial magnetic fields below 40 kG, the rotation rate at low latitudes tends to be slower than the surface rate, contrary to observations. Thus comparison with the observed sunspot rotation rate seem to favor stronger fields as the progenitor of solar active regions.  However, because of the limitations of our model, we recognize that there remain large uncertainties in our results of the azimuthal motion of emerging loops and how they relate to the observed sunspot rotations.  These limitations arise because our simulations stop before the tube enters the solar surface shear layer, and because we do not address the deformation and fragmentation of the flux tubes which may result in a stronger and more complex coupling of the magnetic fields with the fluid motion, especially in the upper layers of the convection zone.

Overall, the results in this study suggest that the initial field strength of active region progenitor flux tubes needs to be sufficiently large, probably $\ge$ 40 kG, in order for them to satisfy the Joy's Law trend for mean tilt angles as well as the observed amount of scatter of the tilt angles about the mean Joy's Law behavior.  Weaker magnetic fields tend to produce too large a scatter to be consistent with the observed results.  Mid-field strengths of 40 kG\,--\,50 kG, which take the longest to rise, do the best job at matching magnetogram observations of Joy's Law dependence.  Although only 50 kG or greater field strengths can rotate at or faster than the solar surface rate.  So, according to our thin flux tube approach, magnetic field values need to be of moderate to large field strengths for tubes with fluxes of $10^{21}$ Mx and $10^{22}$ Mx to produce sunspot rotation behavior.  The addition of multiple magnetic flux values to this study allows us to perform a more comprehensive study on the dependence of flux tube evolution with regard to magnetic flux.  We find, that with convective effects, the tube rise time substantially decreases with decreasing flux.  There also is no statistically significant dependence on the Joy's Law trend in relation to the value of the magnetic flux, although the spread of the tilt angles about the Joy's Law trend does increase as flux decreases.  Convective effects are important to flux tube development at these field strengths and magnetic flux values, and should be incorporated into future studies. 

With this new study, all magnetic field strengths now show a positive Joy's Law trend, and we are now learning that the magnetic field asymmetry of sunspots could be due to a retrograde flow of plasma along the emerging flux tube.  As a result, the estimate of the required $\ge$ 40 kG magnetic field strength is based here mainly on the tilt angle scatter and rotation rates.  However, identifying the rotation rate of the emerging flux tube, in particular, is an aspect where the thin flux tube approach may be lacking in its ability to adequately capture all of the physical processes involved.  We suggest that this aspect of flux tube evolution would be a useful topic for further high-resolution three-dimensional MHD simulations of global flux emergence. In addition, more of these simulations are needed using realistic active region flux tube parameters to further constrain the magnetic field strength generated by the solar dynamo based on observational diagnostics.

%

%
\begin{acks}
This work is supported by NASA SHP grant NNX10AB81G to the National Center for Atmospheric Research (NCAR).  NCAR is sponsored by the National Science Foundation.  We would like to thank Nick Featherstone for reading our manuscript as our internal reviewer, and for offering helpful comments.  Also, we would like to thank our referee for offering a critical review of our manuscript, which contributed to the production of a more substantial article.    
 \end{acks}

%

\end{article} 
\end{document}